# NEAR FORWARD $pp$ ELASTIC SCATTERING AT LHC AND NUCLEON STRUCTURE

M. M. ISLAM* and R. J. LUDDY†

*Department of Physics, University of Connecticut, Storrs, CT 06269, USA*
** islam@phys.uconn.edu*
*† rjluddy@phys.uconn.edu*

A. V. PROKUDIN

*Dipartmento di Fisica Teorica, Università Degli Studi di Torino, Via Pietro Giuria 1, 10125 Torino, Italy*
*Sezione INFN di Torino, Italy*
*Institute for High Energy Physics, 142281 Protvino, Russia*
*prokudin@to.infn.it*

**Abstract**

High energy $pp$ and $\bar{p}p$ elastic scattering carried out at CERN ISR and SPS Collider and at Fermilab Tevatron are studied first in a model where the nucleon has an outer cloud and an inner core. Elastic scattering is viewed as primarily due to two processes: a) diffraction scattering originating from cloud-cloud interaction; b) a hard or large $|t|$ scattering originating from one nucleon core scattering off the other via vector meson $\omega$ exchange, while their outer clouds interact independently. For small $|t|$ diffraction dominates, but the hard scattering takes over as $|t|$ increases. The $\omega$-exchange amplitude shows that $\omega$ behaves like an elementary vector meson at high energy, contrary to a regge pole behavior. This behavior, however, can be understood in the nonlinear $\sigma$-model where $\omega$ couples to a topological baryonic current like a gauge boson, and the nucleon is described as a topological soliton. Further investigation shows that the underlying effective field theory model is a gauged Gell-Mann-Levy type linear $\sigma$-model that has not only the pion sector and the Wess-Zumino-Witten action of the nonlinear $\sigma$-model, but also a quark sector where quarks and antiquarks interact via a scalar field. The scalar field vanishes near the center of the nucleon, but rises sharply at some critical distance to its vacuum value $f_\pi$ leading to a $q\bar{q}$ condensate analogous to a BCS condensate in superconductivity. The nucleon structure that emerges then is that the nucleon has an outer cloud of $q\bar{q}$ condensed ground state, an inner core of topological baryonic charge probed by $\omega$, and a still smaller quark-bag containing massless valence quarks. Large $|t|$ $pp$ elastic scattering is attributed to valence quark-quark elastic scattering, which is taken to be due to the hard pomeron (BFKL pomeron with next to leading order corrections). The parameters in the model are determined by requiring that they satisfactorily describe the known asymptotic behavior of $\sigma_{tot}(s)$ and $\rho(s)$, and the measured $\bar{p}p$ $d\sigma/dt$ at $\sqrt{s}$ = 546 GeV, 630 GeV, and 1.8 TeV. The model is then used to predict $pp$ elastic differential cross section at LHC at $\sqrt{s}$ = 14 TeV and $|t|$ = 0 - 10 GeV$^2$. Our predicted $d\sigma/dt$ at 14 TeV is found to be very different from those predicted by the impact-picture model and the eikonalized pomeron-reggeon model. Precise measurement of $d\sigma/dt$ at LHC by the TOTEM group will be able to distinguish between these models. If our predicted $d\sigma/dt$ is quantitatively confirmed, then it will indicate that various novel ideas developed to describe the nucleon combine and lead to a unique physical description of its structure.

## 1. Introduction

High energy proton-proton ( $pp$ ) and antiproton-proton ( $\bar{p}p$ ) elastic scattering have been measured at CERN Intersecting Storage Rings (ISR)[1] and Super Proton Synchrotron (SPS) Collider,[2,3] and at Fermilab fixed target experiments[4] and Tevatron.[5,6] These experimental studies covering the center of mass (c.m.) energy range 23 - 62 GeV at ISR to 2 TeV at Tevatron will soon reach a new milestone when the Large Hadron Collider (LHC) at CERN comes into operation in 2007. At LHC $pp$ elastic differential cross section is planned to be measured at the unprecedented c.m. energy of 14 TeV by the experimental group TOTEM[7] (acronym for TOTal and Elastic Measurement) in the momentum transfer range $|t| = 0$ to $|t| \simeq 10$ GeV$^2$.

An experimental effort of this magnitude, not surprisingly, has led to many theoretical models. Most of these models have focused on the diffraction region: $|t| \simeq 0 - 0.5$ GeV$^2$; for example, i) single pomeron exchange with a trajectory $\alpha_p(t) = 1.08 + 0.25t$ ;[8] ii) multiple pomeron exchanges with single- and double- diffractive dissociation;[9] iii) incident proton viewed as made up of two color dipoles in the target proton rest frame.[10] Three groups have predicted $pp$ elastic $d\sigma/dt$ at LHC all the way from $|t| = 0$ to $|t| = 10$ GeV$^2$ on the basis of three different models: 1) impact-picture model based on the Cheng-Wu calculations of QCD tower diagrams,[11] 2) eikonalized pomeron-reggeon model using conventional Regge pole approach, but with multiple pomeron, reggeon exchanges included;[12] 3) nucleon-structure model where the nucleon has an outer cloud of quark-antiquark condensed ground state, an inner core of topological baryonic charge, and a still smaller quark-bag of valence quarks.[13,14] A QCD-inspired eikonalized model has also been proposed[15] to predict $pp$ $d\sigma/dt$ at $\sqrt{s}$ =14 TeV for $|t| = 0 - 2.0$ GeV$^2$. This wide array of models attempting to describe $pp$ elastic scattering at LHC reflects the view that quantitative understanding of this process will provide fundamental insight into the nonperturbative and the purturbative QCD dynamics.

Our goals in this article are to provide an overview of our phenomenological investigation that originally began in the late seventies, indicate successive developments of the investigation, describe details of our recent work on large $|t|$ elastic scattering, and present quantitative calculations and LHC results. The paper is organized in several sections. In Sec. 2, we discuss initial developments[16,17] of the investigation that ended with the analysis of ISR and SPS Collider data.[18] In Sec. 3, we show how a gauged linear $\sigma$-model emerges as an underlying effective field theory model leading to our phenomenological description of the nucleon structure.[19,20] Sec. 4 traces the development of an asymptotic description from the phenomenological studies.[21] Sec. 5 details our recent investigation[14] of hard pomeron contribution to $pp$ elastic scattering. Sec. 6 shows how quantitative calculations are carried out and the results we obtain for LHC. Finally, in Sec. 7, we present our concluding remarks. Appendix A provides some relevant mathematical details of Sec. 3 and Appendix B those of Sec. 5.

## 2. Initial Developments

The general framework adopted by us to describe high energy elastic scattering is the Fourier-Bessel or impact parameter representation of the elastic scattering amplitude[22]:

$$T(s,t) = ipW\int_0^\infty bdbJ_0(bq)\Gamma(s,b), \qquad (2.1)$$

where $p$ is the c.m. momentum, $W = \sqrt{s}$ is the c.m. energy, $q = \sqrt{|t|} = 2p\sin\theta/2$, where $\theta$ is the c.m. scattering angle, and $\Gamma(s,b)$ is the profile function, which is related to the eikonal function $\chi(s,b) : \Gamma(s,b) = 1 - \exp[i\chi(s,b)]$. The representation (2.1) holds for all energies and momentum transfers under certain general conditions,[23] but its usefulness lies at high energy, particularly because of the semi-classical approximation

$$e^{2i\delta_l(s)} \simeq e^{i\chi(s,b)}\Big|_{b=\frac{l+1/2}{p}} \qquad (2.2)$$

and its connection with unitarity. Here, $\delta_l(s)$ is the partial wave phase shift. Since unitarity implies $\left|\exp[2i\delta_l(s)]\right| \le 1$ and $\mathrm{Im}\,\delta_l(s) \ge 0$, it follows from (2.2) that at high energy $\left|\exp[i\chi(s,b)]\right| \le 1$, and if we write $\exp[i\chi(s,b)] = \eta(s,b)\exp[i\chi_R(s,b)]$, then $\eta(s,b) = \exp[-\chi_I(s,b)]$ satisfies the requirement $0 \le \eta(s,b) \le 1$ $(\chi = \chi_R + i\chi_I)$.

The elastic differential cross section is given by

$$\frac{d\sigma}{dt} = \frac{\pi}{p^2 s}\left|T(s,t)\right|^2, \qquad (2.3)$$

and the total cross section is given by the optical theorem:

$$\mathrm{Im}\,T(s,0) = \frac{pW}{4\pi}\sigma_{tot}(s). \qquad (2.4)$$

Initial phenomenological investigations[16,17] led us to the following description. The nucleon has an outer cloud and an inner core (Fig. 1). High energy elastic scattering is primarily due to two processes (Fig. 2a,b): a) a glancing collision where the outer cloud of one nucleon interacts with that of the other giving rise to diffraction scattering; b) a hard, or large $|t|$ collision where one nucleon core scatters off the other core via vector meson $\omega$ exchange, while their outer clouds overlap and interact independently. In the small $|t|$ region diffraction dominates, but the hard scattering takes over as $|t|$ increases. We took the elastic amplitude as the sum of two amplitudes: $T(s,t) = T_D(s,t) + T_H^\omega(s,t)$, where $T_D(s,t)$ is the diffraction amplitude, and $T_H^\omega(s,t)$ is the hard scattering amplitude due to $\omega$-exchange. For $\bar{p}p$ and $pp$ scattering

$$T_D^{\bar{p}p}(s,t) = T_D^+(s,t) + T_D^-(s,t), \qquad (2.5a)$$

$$T_D^{pp}(s,t) = T_D^+(s,t) - T_D^-(s,t), \qquad (2.5b)$$

where $T_D^+, T_D^-$ are the crossing even and crossing odd amplitudes. In the range of momentum transfer where diffraction dominates, we regard the diffraction amplitude to be primarily $T_D^+(s,t)$ and therefore the same for $\bar{p}p$ and $pp$.

$T_D^+(s,t)$ can also be expressed as a Fourier-Bessel transform:

$$T_D^+(s,t) = ipW\int_0^\infty b\,db\,J_0(bq)\Gamma_D^+(s,b)\,. \tag{2.6}$$

We choose $\Gamma_D^+(s,b)$ to be an even Fermi profile function[18]:

$$\Gamma_D^+(s,b) = g(s)\left[\frac{1}{1+e^{\frac{b-R}{a}}} + \frac{1}{1+e^{-\frac{b+R}{a}}} - 1\right] \tag{2.7}$$

with $R$ and $a$ energy dependent:

$$R = R(s) = R_0 + R_1(\ln s - i\tfrac{\pi}{2}) \tag{2.8a}$$

$$a = a(s) = a_0 + a_1(\ln s - i\tfrac{\pi}{2})\,. \tag{2.8b}$$

$g(s)$ is a crossing even complex energy dependent function: $g^*(se^{i\pi}) = g(s)$ that asymptotically becomes a real positive constant. $\Gamma_D^+(s,b)$ in (2.7) can be expressed as

$$\Gamma_D^+(s,b) = g(s)\frac{\sinh(R/a)}{\cosh(R/a)+\cosh(b/a)}\,, \tag{2.9}$$

which shows that for $b >> a$ and $R$, it falls off exponentially

$$\Gamma_D^+(s,b) \simeq 2g(s)\sinh(R/a)e^{-\frac{b}{a}}\,, \qquad (\tfrac{b}{a} \to \infty)\,. \tag{2.10}$$

Hence the integral in (2.6) with $\Gamma_D^+(s,b)$ given by (2.7) converges even when $q = 0$ leading to a finite forward amplitude.

When $q \neq 0$, the integral in (2.6) can be done analytically by noticing that the poles of (2.7) in the right-hand $b$-plane are at $b = R \pm i\pi(2n+1)$.[16] If $q$ is large, the dominant contribution corresponds to $n = 0$, and we obtain[18]

$$T_D^+(s,t) \simeq ipWag(s)\left\{-i\pi[R+i\pi a]H_0^{(1)}[q(R+i\pi a)] + i\pi[R-i\pi a]H_0^{(2)}[q(R-i\pi a)]\right\} \tag{2.11}$$

which is a crossing even amplitude.

The hard scattering amplitude due to $\omega$ exchange (Fig. 2b) turns out to be

$$T_1^\omega(s,t) = \pm\tilde{\gamma}\exp[i\hat{\chi}(s,0)]\,\frac{sF^2(t)}{m_\omega^2 - t}\,, \tag{2.12}$$

where the + (–) sign corresponds to $\bar{p}p$ ( $pp$ ) scattering and $\tilde{\gamma}$ is a positive real constant. $\exp[i\hat{\chi}(s,0)]$ originates from the accompanying cloud-cloud interaction. The factor of $s$ originates from spin 1 of $\omega$. The t-dependence is the product of two form factors and the $\omega$ propagator, which shows that $\omega$ probes two density distributions corresponding to the two nucleon cores. Eq. (2.12) was first obtained using a simple model based on a non-planar Feynman diagram.[24] Writing $\exp[i\hat{\chi}(s,0)] = \hat{\eta}(s)\exp[i\hat{\theta}(s)]$ and $\tilde{\gamma}\hat{\eta}(s) = \hat{\gamma}(s)$, Eq. (2.12) takes the form

$$T_1^\omega(s,t) = \pm\hat{\gamma}(s)\exp[i\hat{\theta}(s)]\,\frac{sF^2(t)}{m_\omega^2 - t}\,. \tag{2.13}$$

The form factors in Eq. (2.13) were initially determined by requiring that $T_1^\omega(s,t)$ has to fall off exponentially in $q = \sqrt{|t|}$. This requirement follows from an observation by Orear[25] that large $|t|$ $pp$ $d\sigma/dt$ falls off exponentially in transverse momentum $p_\perp = p\sin\theta \simeq q$. Such a fall-off occurs if the two nucleons interact via a smoothed Yukawa potential $V(r) \sim \dfrac{\exp[-m(r^2+\beta^2)^{\frac{1}{2}}]}{(r^2+\beta^2)^{\frac{1}{2}}}$, $(m = m_\omega)$, which leads to the form factors[16]

$$F^2(t) = \beta(m^2-t)^{\frac{1}{2}} K_1[\beta(m^2-t)^{\frac{1}{2}}], \qquad (2.14)$$

where $K_1(z)$ is the modified Bessel function. Since for |z| large, $K_1(z) \sim e^{-z}$, we find $T_1^\omega(s,t)$ in Eq. (2.13) to behave as $e^{-\beta q}$ for $|t|$ large and to lead to an Orear fall-off. $\beta$ here is a length scale that determines the size of the nucleon core.

In the calculation of the full amplitude, $T_1^\omega(s,t)$ (due to a single hard collision) has to be multiplied by an absorption factor $[1-\Gamma_D^{\bar{p}p}(s,0)]$, or $[1-\Gamma_D^{pp}(s,0)]$ due to diffraction. A crossing odd contribution at zero impact parameter can occur in the absorption factor, so that we have to consider $\Gamma_D^{\bar{p}p}(s,0) = \Gamma_D^+(s,0) + \Gamma_D^-(s,0)$ and $\Gamma_D^{pp}(s,0) = \Gamma_D^+(s,0) - \Gamma_D^-(s,0)$.

Using the quantitative formulation outlined above, $pp$ elastic data from ISR[1] at $\sqrt{s}$ = 23.5 GeV and 53 GeV and $\bar{p}p$ elastic data from SPS Collider[2] at $\sqrt{s}$ = 546 GeV were analyzed.[18] There were four energy independent parameters: $R_0, R_1, a_0, a_1$ and six energy dependent parameters: $|g(s)|, \arg g(s), \hat{\gamma}(s), \hat{\theta}(s), \mathrm{Re}\,\Gamma_D^-(s), \mathrm{Im}\,\Gamma_D^-(s)$. $\beta$ and $m_\omega$ were kept fixed at values $\beta = 3.075$ GeV$^{-1}$ and $m_\omega = 0.801$ GeV found earlier.[16] A satisfactory quantitative description of $pp$ and $\bar{p}p$ elastic scattering was obtained[18] indicating that the structure of the nucleon having an outer cloud and an inner core probed by $\omega$ appears to be supported at this point.

## 3. Gauged Linear $\sigma$-Model as Underlying Effective Field Theory Model

In the context of our above analysis of ISR and SPS collider data, a critical question was raised: namely, our $\omega$-exchange amplitude, Eq. (2.12), shows that $\omega$ behaves as an elementary vector meson. This can be seen from the $t$-dependence of the amplitude, which is that of a meson probing two density distributions, and more importantly, from its $s$-dependence – the factor of $s$ that corresponds to spin 1 of an elementary vector meson. On the other hand, in the widely used Regge pole model of high energy elastic scattering, $\omega$ is treated as a Regge pole, where the $s$-dependence of the $\omega$-exchange amplitude is $s^{\alpha_\omega(t)}$, $\alpha_\omega(t)$ being the $\omega$ trajectory: $\alpha_\omega(t) = 0.44 + 0.92t$.[8] This implies for momentum transfer $|t|$ of the order of 1 GeV$^2$, in contrast to our calculations, the $\omega$-exchange amplitude should be negligible compared to the diffraction amplitude.

An answer to this question came from the nonlinear $\sigma$-model where $\omega$ couples to the baryonic current and the baryonic current is topological, i.e. geometrical in origin.[26] What this model states is that – it is an effective field theory model, and, as long as it holds, the baryonic current continues to behave as topological current and $\omega$ coupled to this current as a gauge boson continues to behave as a gauge boson, that is, as an elementary vector meson. And we seem to be seeing this behavior. Fortunately, this idea could be tested. From our $\omega NN$ form factor $F(t)$ given by Eq. (2.14), the baryonic charge distribution probed by $\omega$ could be obtained by taking the Fourier transform, and then, from the charge distribution the classical pion field (pion profile function) which gives rise to it could be calculated. This then could be compared with the pion profile function obtained from the nonlinear $\sigma$-model that describes the nucleon as a topological soliton, or skyrmion.[26] The results of this investigation are shown in Fig. 3.[27] The solid line represents the pion profile function obtained from our form factor. The dotted and the dashed lines represent the pion profile functions obtained by Meissner, Kaiser, and Weise[28] in their minimal (dotted line) and complete (dashed line) soliton models. As can be seen, our curve is quite compatible with theirs given that our calculations are from the domain of c.m. energy 23 GeV and larger, whereas their calculations are from the energy region of order 1 GeV – two totally different regimes of physics. Also, the root-mean-square (r.m.s.) radius of the soliton obtained by us is 0.44 F while that by Meissner et al. is approximately 0.5 F.

Another interesting question was raised at this point. If $\omega$ is that important in our calculations, then why not $\rho$ and $a_1$? After all, in low energy $NN$ interactions it is known that besides $\omega$, the vector mesons $\rho$ and $a_1$ play important roles. The answer to this question is again provided by the nonlinear $\sigma$-model. What this model states is that we can introduce $\rho, a_1$, and $\omega$ – all on the same footing as gauge bosons. While $\rho$ and $a_1$ couple via isospin currents, $\omega$ couples via baryonic current.[29] The isospin currents are Noether currents, which are the usual currents occurring in perturbative field theory. $\rho$ and $a_1$ therefore should be amenable to perturbative calculations and should reggeize. Consequently, their contributions in high energy elastic scattering would be negligible. On the other hand, $\omega$ couples to the baryonic current and the baryonic current is topological. It is conserved because of its geometrical structure, unlike the Noether currents which are conserved because of equations of motion. $\omega$, therefore, evades reggeization because of its non-perturbative coupling to the baryonic current.

We now face a fundamental question: If our investigation of high energy elastic scattering provides evidence that the nucleon is a topological soliton described by the nonlinear $\sigma$-model, then how do we understand the large soliton mass consistently predicted by the nonlinear $\sigma$-model (typically, $m_{soliton} \sim 1500$ MeV while $m_N = 939$ MeV)? The large mass problem[29,30] has been a central difficulty of chiral soliton models, and has been extensively studied by Zhang and Mukhopadhyay.[31] They found that if the axial vector coupling $g_A$ of the nucleon is close to the experimental value, then the soliton mass turns out to be very large (more than 1.6 GeV). This suggests that we have to look beyond the nonlinear $\sigma$-model. To this end, we examine the gauged linear $\sigma$-model and as a first step consider the Gell-Mann-Levy linear $\sigma$-model.[26,32]

The linear $\sigma$-model of Gell-Mann and Levy has $SU(2)_L \times SU(2)_R$ global symmetry and spontaneous breakdown of chiral symmetry. The model is given by the Lagrangian density

$$\mathcal{L} = \bar{\psi}\, i\gamma^\mu \partial_\mu \psi + \tfrac{1}{2}(\partial_\mu \sigma\, \partial^\mu \sigma + \partial_\mu \vec{\pi} \cdot \partial^\mu \vec{\pi}) - G\bar{\psi}[\sigma + i\,\vec{\tau} \cdot \vec{\pi}\, \gamma^5]\psi - \lambda(\sigma^2 + \vec{\pi}^2 - f_\pi^2)^2 . \quad (3.1)$$

In our study, the fermions are identified with quarks, and we start with a two flavor model for quarks (u and d). We introduce a scalar-isoscalar field $\zeta(x)$ and a unitary field $U(x)$ in the following way:

$$\sigma(x) + i\,\vec{\tau} \cdot \vec{\pi}(x) = \zeta(x) U(x) . \quad (3.2)$$

$\zeta(x)$ is the magnitude of the fields $\sigma(x)$ and $\vec{\pi}(x)$: $\zeta(x)=\sqrt{\sigma^2(x)+\vec{\pi}^2(x)}$. Using right and left fermion fields:

$$\psi_R(x)=\tfrac{1}{2}(1+\gamma^5)\psi(x), \qquad \psi_L(x)=\tfrac{1}{2}(1-\gamma^5)\psi(x), \tag{3.3}$$

the Lagrangian density (3.1) can now be written as:

$$\mathcal{L}=\overline{\psi}_R\, i\gamma_\mu\partial_\mu\psi_R+\overline{\psi}_L\, i\gamma^\mu\partial_\mu\psi_L+\tfrac{1}{2}\partial_\mu\zeta\,\partial^\mu\zeta+\tfrac{1}{4}\zeta^2\, tr[\partial_\mu U\,\partial^\mu U^\dagger]$$
$$-G\zeta[\overline{\psi}_L U\psi_R+\overline{\psi}_R U^\dagger\psi_L]-\lambda(\zeta^2-f_\pi^2)^2. \tag{3.4}$$

Under global right and left transformations

$$\psi_R(x)\to R\psi_R(x), \quad \psi_L(x)\to L\psi_L(x), \tag{3.5a}$$
$$U(x)\to LU(x)R^\dagger, \quad \zeta(x)\to\zeta(x), \tag{3.5b}$$

and the Lagrangian density (3.4) is manifestly invariant under these global transformations. If we now consider chiral symmetry ($\psi\to e^{\theta\gamma^5}\psi$), then

$$\psi_R(x)\to e^{\theta}\psi_R(x), \quad \psi_L(x)\to e^{-\theta}\psi_L(x), \tag{3.6a}$$
$$U(x)\to e^{-\theta}U(x)e^{-\theta}, \quad \zeta(x)\to\zeta(x), \tag{3.6b}$$

where $\theta=-iT^a\theta^a$ $\quad(T^a=\tfrac{1}{2}\tau^a)$ and $\theta^a$'s are global. Eq. (3.2) and the transformation (3.6b) show that, if $\zeta(x)$ has a nonvanishing vacuum value $\zeta_0$, then the vacuum value $U_0$ of $U(x)$ changes under chiral symmetry:

$$U_0\to e^{-\theta}U_0\,e^{-\theta} \tag{3.7}$$

indicating spontaneous breakdown of the global chiral symmetry. Writing $U(x)=\exp[i\vec{\tau}\cdot\vec{\varphi}(x)/f_\pi]$, we can identify $\vec{\varphi}(x)$ as the isovector pion field, $f_\pi$ as the pion decay coupling constant ($f_\pi\simeq 93$ MeV), and the pions as Goldstone bosons. When $\zeta(x)$ is replaced by its vacuum value $f_\pi$, the Lagrangian (3.4) represents a nonlinear $\sigma$-model.

Following Bando et al.,[33] we now introduce the idea of a hidden local symmetry. Bando et al. took the symmetry to be $[SU(2)]_{hidden}$. We consider an extended version of this approach, and following Meissner et al.[34] take the symmetry to be $[SU(2)\times U(1)]_{hidden}$. To implement the idea of a hidden local symmetry, one writes $U(x)=\xi_L^\dagger(x)\xi_R(x)$, where $\xi_L(x)$ and $\xi_R(x)$ are $SU(2)$-valued fields which transform in the following way under $[SU(2)_L\times SU(2)_R]_{global}\times[SU(2)\times U(1)]_{local}$:

$$\xi_R(x)\to h(x)\xi_R(x)R^\dagger, \qquad \xi_L(x)\to h(x)\xi_L(x)L^\dagger, \tag{3.8}$$

where $h(x)\in[SU(2)\times U(1)]_{local}$. Therefore,

$$U(x)=\xi_L^\dagger(x)\xi_R(x)\to L\,U(x)R^\dagger \tag{3.9}$$

as required by the global symmetry of the Lagrangian (3.4).

Let us focus on the pion sector of the Lagrangian (3.4):

$$\mathcal{L}_\pi=\tfrac{1}{4}\zeta^2\, tr[\partial_\mu U\,\partial^\mu U^\dagger]$$
$$=-\tfrac{1}{4}\zeta^2\, tr[\partial_\mu\xi_L\,\xi_L^\dagger-\partial_\mu\xi_R\xi_R^\dagger]^2. \tag{3.10}$$

We gauge this symmetry by introducing the vector mesons $\vec{\rho}$ and $\omega$ as gauge bosons of the $[SU(2)\times U(1)]_{hidden}$ symmetry. Gauging is done easily by replacing the ordinary derivative by the covariant derivative: $\partial_\mu \to D_\mu = \partial_\mu + \mathcal{V}_\mu$, where

$$\mathcal{V}_\mu = -\tfrac{i}{2} g[\vec{\tau}\cdot\vec{\rho}_\mu + \omega_\mu] \tag{3.11}$$

and $\mathcal{V}_\mu$ transforms under the hidden symmetry in the following way:

$$\mathcal{V}_\mu \to h\mathcal{V}_\mu h^\dagger + h\partial_\mu h^\dagger . \tag{3.12}$$

The Lagrangian density of the gauged hidden-symmetry model is taken as[33,34]

$$\begin{aligned}\mathcal{L}_\pi &= -\tfrac{1}{4}\zeta^2\, tr[(D_\mu\,\xi_L)\,\xi_L^\dagger - (D_\mu\,\xi_R)\xi_R^\dagger]^2 \\ &\quad -\tfrac{1}{2}\zeta^2\, tr[(D_\mu\,\xi_L)\,\xi_L^\dagger + (D_\mu\,\xi_R)\,\xi_R^\dagger]^2 + \frac{1}{2g^2} tr[F_{\mu\nu}F^{\mu\nu}],\end{aligned} \tag{3.13}$$

where the first term is exactly the same as the original term $\tfrac{1}{4}\zeta^2\, tr[\partial_\mu U\,\partial^\mu U^\dagger]$. The second term is a gauge invariant term that generates the masses of the vector mesons. The third term is the Lagrangian density of the gauge field $\mathcal{V}_\mu(x)$, and $F_{\mu\nu}$ is the nonabelian field tensor: $F_{\mu\nu} = \partial_\mu\mathcal{V}_\nu - \partial_\nu\mathcal{V}_\mu + [\mathcal{V}_\mu,\mathcal{V}_\nu]$. The third or last term in Eq. (3.13) produces a term like $tr[L_\mu, L_\nu]^2$, where $L_\mu = U^\dagger\,\partial_\mu U$.[26] This is the fourth-order derivative term introduced by Skyrme.

We observe at this point that the gauge field $\mathcal{V}_\mu(x)$ of the hidden gauge symmetry gives rise to a left gauge field $A_\mu^L(x)$ and a right gauge field $A_\mu^R(x)$ associated with local left and local right transformations. This can be seen in the following way. Let us consider

$$A_\mu^L(x) = \xi_L^+(x)\mathcal{V}_\mu(x)\,\xi_L(x) + \xi_L^+(x)\partial_\mu\,\xi_L(x) . \tag{3.14}$$

Under simultaneous local $SU(2)_L$ and local hidden gauge transformations,

$$\xi_L(x) \to \xi_L'(x) = h(x)\,\xi_L(x)\,L^\dagger(x)\,, \tag{3.15a}$$

$$\mathcal{V}_\mu(x) \to \mathcal{V}_\mu'(x) = h(x)\mathcal{V}_\mu(x)\,h^\dagger(x) + h(x)\partial_\mu h^\dagger(x)\,, \tag{3.15b}$$

$$A_\mu^L(x) \to A_\mu'^L(x) = L(x)\,A_\mu^L(x)\,L^\dagger(x) + L(x)\partial_\mu\,L^\dagger(x)\,. \tag{3.15c}$$

Eq. (3.15c) shows $A_\mu^L(x)$ indeed transforms as a left gauge field. Proceeding in a similar way, we define

$$A_\mu^R(x) = \xi_R^\dagger(x)\mathcal{V}_\mu(x)\,\xi_R(x) + \xi_R^\dagger(x)\partial_\mu\xi_R(x)\,. \tag{3.16}$$

Under simultaneous local $SU(2)_R$ and local hidden gauge transformations,

$$\xi_R(x) \to \xi_R'(x) = h(x)\,\xi_R(x)\,R^\dagger(x)\,, \tag{3.17a}$$

$$\mathcal{V}_\mu(x) \to \mathcal{V}_\mu'(x) = h(x)\mathcal{V}_\mu(x)\,h^\dagger(x) + h(x)\partial_\mu h^\dagger(x)\,, \tag{3.17b}$$

$$A_\mu^R(x) \to A_\mu'^R(x) = R(x)\,A_\mu^R(x)\,R^\dagger(x) + R(x)\partial_\mu R^\dagger(x)\,, \tag{3.17c}$$

which shows $A_\mu^R(x)$ transforms as a right gauge field.

The above development shows that the gauge field $\mathcal{V}_\mu(x)$ of hidden symmetry induces left and right gauge fields through which it can interact with left and right quark fields in a gauge invariant manner. We now include these interactions and write the Lagrangian for the quark sector as

$$\mathcal{L}_q = \overline{\psi}_L(x)\gamma^\mu(\partial_\mu + A^L_\mu(x))\psi_L(x) + \overline{\psi}_R(x)\gamma^\mu(\partial_\mu + A^R_\mu(x))\psi_R(x)\,. \tag{3.18}$$

The gauge invariance of $\mathcal{L}_q$ under local left and right transformations is manifest, since $\psi_L(x) \to \psi'_L(x) = L(x)\psi_L(x),\quad \psi_R(x) \to \psi'_R(x) = R(x)\psi_R(x)$, and $A^L_\mu(x),\, A^R_\mu(x)$ transform like gauge bosons (Eqs. (3.15c) and (3.17c)). The interaction of quarks with $\zeta(x)$ and $U(x)$ fields in Eq. (3.4) will be included later on. The gauge fields $A^L_\mu(x)$ and $A^R_\mu(x)$ act as external gauge fields in $\mathcal{L}_q$, and we now focus on this Lagrangian density.

To take into account quantum effects in the quark sector, which lead to anomalous actions, we consider the action functionals, or effective actions $W[A^L(x)]$ and $W[A^R(x)]$ of the gauge fields $A^L_\mu(x)$ and $A^R_\mu(x)$:

$$e^{iW[A^L(x)]} = \frac{1}{\mathcal{N}_L}\int d\psi_L d\overline{\psi}_L\, e^{\int d^4x\,\overline{\psi}_L(x)\gamma^\mu(\partial_\mu + A^L_\mu(x))\psi_L(x)}\,, \tag{3.19}$$

and

$$e^{iW[A^R(x)]} = \frac{1}{\mathcal{N}_R}\int d\psi_R d\overline{\psi}_R\, e^{\int d^4x\,\overline{\psi}_R(x)\gamma^\mu(\partial_\mu + A^R_\mu(x))\psi_R(x)}\,. \tag{3.20}$$

The normalizations $\mathcal{N}_L$ and $\mathcal{N}_R$ are chosen such that the free field terms are subtracted out. Here, $A^L_\mu(x) = -iT^a A^{La}_\mu(x),\, A^R_\mu(x) = -iT^a A^{Ra}_\mu(x)$ and the $T^a$'s are the generators of the symmetry group. For the present discussion, we take a three flavor model[35] for quarks (u, d, s), so that the symmetry group under consideration is local $SU(3)_L \times SU(3)_R$. The generators $T^a$'s are $T^a = \frac{1}{2}\lambda^a$, where $\lambda^a$'s are the Gell-Mann matrices: $[T^a, T^b] = i\, f^{abc}T^c$, $tr\,[T^aT^b] = \frac{1}{2}\delta^{ab}$. The integrals over $d^4x$ in Eqs. (3.19) and (3.20) are in Euclidean space: $x^0 \to -i\,x^4$, $x = (x^1, x^2, x^3, x^4)$. $\gamma^\mu$'s are antihermitian: $(\gamma^\mu)^\dagger = -\gamma^\mu$ $(\mu = 1,2,3,4)$, $\gamma^0 = -i\,\gamma^4$, $\gamma^5 = i\,\gamma^0\gamma^1\gamma^2\gamma^3 = \gamma^4\gamma^1\gamma^2\gamma^3$; $\overline{\psi}_L = \psi^\dagger_L\gamma^4 = \overline{\psi}\, a_+$, $\overline{\psi}_R = \psi^\dagger_R\gamma^4 = \overline{\psi}\, a_-$, and $a_+ = \frac{1}{2}(1+\gamma^5)$, $a_- = \frac{1}{2}(1-\gamma^5)$. The Euclidean metric is $g_{11} = g_{22} = g_{33} = g_{44} = -1$.

To derive the anomalous actions of the action functionals $W[A^L(x)]$ and $W[A^R(x)]$ originating from the gauge dependence of the fermion measure, we follow the strategy of extending the gauge field to a fifth dimension (a parameter space), determine the variation of the effective action due to an infinitesimal change of the parameter, and then integrate all such variations to obtain the full anomalous action. The details of carrying out this program following the earlier work[36] are given in appendix A. The final result is

$$e^{i\left(W[A^L(x)+W[A^R(x)]\right)} = e^{i\Gamma_{WZW}[U(x),\mathcal{V}(x)]}\; e^{i\left(W_L[\mathcal{V}(x)]+W_R[\mathcal{V}(x)]\right)} \tag{3.21}$$

where

$$i\,\Gamma_{WZW}[U(x),\mathcal{V}(x)] = 2\int_0^1 dt\int d^4x\; tr_G[U^\dagger(x,t)\,\partial_t U(x,t)a_L(\mathcal{A}^L) + U(x,t)\,\partial_t U^\dagger(x,t)a_R(\mathcal{A}^R)]\,. \tag{3.22}$$

As the parameter $t$ varies from 0 to 1, $U(x,t)$ interpolates the field $\mathscr{A}_\mu^L(x,t)$ between $\mathcal{V}_\mu(x)$ and $A_\mu^L(x)$, and $U^\dagger(x,t)$ interpolates the field $\mathscr{A}_\mu^R(x,t)$ between $\mathcal{V}_\mu(x)$ and $A_\mu^R(x)$ ($U(x,1)=\xi^2(x)=U(x), \quad U(x,0)=\xi(x), \quad \xi_L^\dagger(x)=\xi_R(x)=\xi(x)$). $a_L(\mathscr{A}^L)$ and $a_R(\mathscr{A}^R)$ are the nonabelian anomalies of the left and right currents: $a_L(\mathscr{A}^L)=D_\mu(\mathscr{A}^L)\, j_L^\mu(x,t)$, $a_R(\mathscr{A}^R)=D_\mu(\mathscr{A}^R)\, j_R^\mu(x,t)$. Their explicit forms are given in Eqs. (A.20) and (A.34). The effective action $W_L[\mathcal{V}(x)]$ is given by

$$e^{iW_L[\mathcal{V}(x)]}=\frac{1}{\mathscr{N}_L}\int d\psi_L\; d\overline{\psi}_L\; e^{\int d^4x\overline{\psi}_L(x)\gamma^\mu(\partial_\mu+\mathcal{V}_\mu(x))\psi_L(x)}\,. \tag{3.23}$$

$W_L[\mathcal{V}(x)]$ is invariant under left gauge transformations, since it only involves the gauge field $\mathcal{V}_\mu(x)$ of the hidden symmetry and does not respond to local left transformations. This statement can be made manifest at the level of the action by writing

$$e^{iW_L[\mathcal{V}(x)]}=\frac{1}{\mathscr{N}_L}\int d\psi_L^0\; d\overline{\psi}_L^0\; e^{\int d^4x\overline{\psi}_L^0(x)\gamma^\mu(\partial_\mu+\mathcal{V}_\mu(x))\psi_L^0(x)}\,, \tag{3.24}$$

and taking $\psi_L^0(x)=\xi_L(x)\psi_L(x)$. Because $\xi_L(x)\to h(x)\xi_L(x)L^\dagger(x)$ and $\psi_L(x)\to L(x)\psi_L(x)$, $\psi_L^0(x)$ only responds to the hidden symmetry: $\psi_L^0(x)\to h(x)\psi_L^0(x)$. We follow the same reasoning for the effective action $W_R[\mathcal{V}(x)]$ given by

$$e^{iW_R[\mathcal{V}(x)]}=\frac{1}{\mathscr{N}_R}\int d\psi_R\; d\overline{\psi}_R\; e^{\int d^4x\overline{\psi}_R(x)\gamma^\mu(\partial_\mu+\mathcal{V}_\mu(x))\psi_R(x)}\,, \tag{3.25}$$

and write

$$e^{iW_R[\mathcal{V}(x)]}=\frac{1}{\mathscr{N}_R}\int d\psi_R^0\; d\overline{\psi}_R^0\; e^{\int d^4x\overline{\psi}_R^0(x)\gamma^\mu(\partial_\mu+\mathcal{V}_\mu(x))\psi_R^0(x)}\,, \tag{3.26}$$

where $\psi_R^0(x)=\xi_R(x)\psi_R(x)$. Again, as $\xi_R(x)\to h(x)\xi_R(x)R^\dagger(x)$ and $\psi_R(x)\to R(x)\psi_R(x)$, $\psi_R^0(x)$ only responds to the hidden symmetry: $\psi_R^0(x)\to h(x)\psi_R^0(x)$, and the action in (3.26) is manifestly invariant under local right transformations.

We can now combine (3.24) and (3.26) and rewrite (3.21) in the form

$$e^{i\left(W[A^L(x)]+W[A^R(x)]\right)}=e^{i\Gamma_{WZW}[U(x),\mathcal{V}(x)]}\,\frac{1}{\mathscr{N}}\int d\psi^0\, d\overline{\psi}^0\, e^{\int d^4x\overline{\psi}^0(x)\gamma^\mu(\partial_\mu+\mathcal{V}_\mu(x))\psi^0(x)}\,, \tag{3.27}$$

where $\psi^0(x)=\psi_L^0(x)+\psi_R^0(x)$ and $\mathscr{N}=\mathscr{N}_L\,\mathscr{N}_R$. The action in the quark sector in (3.27) shows that we can easily include the quark-antiquark interaction occurring in (3.4) by writing it in the form $G\overline{\psi}^0\psi^0$. In fact, we can write an effective action for the fields $U$, $\zeta$ and $\psi^0$ on the basis of (3.4), (3.13) and (3.27) with $\mathcal{V}_\mu(x)$ treated as an external gauge field:

$$e^{iW[U,\mathcal{V},\zeta,\psi^0]}=e^{i\Gamma_{WZW}[U,\mathcal{V}]}\,\frac{1}{N}\int dU\, d\zeta\, d\psi^0\, d\overline{\psi}^0\; e^{iS[U,\zeta,\mathcal{V}]+iS[\zeta,\psi^0,\mathcal{V}]}\,. \tag{3.28}$$

Here, $S[U,\zeta,\mathcal{V}]$ is the action in the pion sector:

$$S[U,\zeta,\mathcal{V}]=\int d^4x\,\mathcal{L}_\pi(U,\zeta,\mathcal{V})\,, \tag{3.29}$$

with $\mathcal{L}_\pi$ given by (3.13) in the unitary gauge: $\xi_L^\dagger(x)=\xi_R(x)=\xi(x)$, $U(x)=\xi^2(x)$. $S[\zeta,\psi^0,\mathcal{V}]$ is the action in the quark-scalar sector:

$$S[\zeta,\psi^0,\mathcal{V}] = \int d^4x\,[\overline{\psi}^0\, i\,\gamma^\mu\,(\partial_\mu + \mathcal{V}_\mu)\psi^0 + \tfrac{1}{2}\partial_\mu\zeta\,\partial^\mu\zeta - G\,\zeta\,\overline{\psi}^0\psi^0 - \lambda(\zeta^2 - f_\pi^2)^2]\,, \quad (3.30)$$

written in Minkowski space. $N$ is an appropriate normalization constant. If we approximate the scalar field $\zeta(x)$ by its vacuum value $f_\pi$ in the pion sector, we obtain

$$e^{iW[U,\mathcal{V},\zeta,\psi^0]} \simeq e^{i\Gamma_{WZW}[U,\mathcal{V}]}\,\frac{1}{N}\int dU\, e^{iS[U,f_\pi,\mathcal{V}]}\int d\zeta\, d\psi^0\, d\overline{\psi}^0\, e^{iS[\zeta,\psi^0,\mathcal{V}]}\,, \quad (3.31)$$

where

$$S[U,f_\pi,\mathcal{V}] = \int d^4x\,\Big\{\tfrac{1}{4}f_\pi^2\, tr[\partial_\mu U\,\partial^\mu U^\dagger] - \tfrac{1}{2}f_\pi^2\, tr[(D_\mu\xi^\dagger)\xi + (D_\mu\xi)\xi^\dagger]^2 + \frac{1}{2g^2}tr[F_{\mu\nu}F^{\mu\nu}]\Big\}\,. \quad (3.32)$$

The topological solitons of the nonlinear $\sigma$-model ($NL\sigma M$) are described by the combined action: $S[U,f_\pi,\mathcal{V}] + \Gamma_{WZW}[U,\mathcal{V}]$.[28]

The Wess-Zumino-Witten action in its simplest approximation[34] can be written as $\Gamma_{WZW}[U,\mathcal{V}] = \int d^4x\,\mathcal{L}_{WZW}$, where

$$\mathcal{L}_{WZW} = g_\omega\,\omega_\mu\,B^\mu\,, \quad (3.33)$$

and $B^\mu$ is the conserved topological baryonic current

$$B^\mu(x) = \frac{1}{24\pi^2}\epsilon^{\mu\nu\rho\sigma}\, tr[U^\dagger\,\partial_\nu U\,U^\dagger\,\partial_\rho U\,U^\dagger\,\partial_\sigma U]\,, \quad (3.34)$$

and $U(x) = \exp[i\,\vec{\tau}\cdot\vec{\varphi}(x)/f_\pi]$. Eq. (3.33) shows that the vector meson $\omega$, indeed, couples to the baryonic current just like a gauge boson. We note that the masses of $\rho$ and $\omega$ are generated by the second term on the RHS of (3.32) and leads to the KSFR relation $m_\rho^2 = m_\omega^2 = 2\,f_\pi^2 g^2$. The last term in (3.32), as we noted earlier, reproduces Skyrme's fourth order derivative term.

We notice in (3.31) we have a quark-scalar sector described by the action $S[\zeta,\psi^0,\mathcal{V}]$, which is not present in the $NL\sigma M$. The reasons are: first, in the $NL\sigma M$ the scalar field $\zeta(x)$ is replaced by its vacuum value $f_\pi$ from the very beginning; so, the scalar degree of freedom never appears; second, it is implicitly assumed that once the WZW action originating from the quark sector is taken into account, no further interactions from the quark sector need be considered. These, of course, mean in realistic terms, the quark-scalar sector is taken to be a non-interacting Dirac sea of quarks and antiquarks, which does not contribute to the total energy. On the other hand, from $S[\zeta,\psi^0,\mathcal{V}]$ in (3.30), we find that the scalar field $\zeta(x)$ provides an important interaction between quarks and antiquarks by making them massive. It, therefore, indicates the presence of an interacting Dirac sea. What one finds is that – if the scalar field $\zeta$ has a critical behavior in the static case as shown in Fig. 4, that is, $\zeta = 0$ for $r$ less than a critical distance $r_c$ and rising sharply at $r = r_c$ to its vacuum value $f_\pi$, then the interacting Dirac sea has considerably less energy than the non-interacting Dirac sea.[19] This is a condensation phenomenon where, by acquiring mass, the quarks reduce the total energy of the ground state analogous to the BCS condensation in superconductivity. If $\zeta(r)$ is taken to be $f_\pi\,\theta(r - r_c)$, then the ground state energy is reduced by an infinite amount.[19] Obviously, by taking a sufficiently sharp behavior of $\zeta(r)$, the ground state energy can be reduced by 500 GeV. This would resolve the large soliton mass problem of the topological soliton model mentioned earlier. Furthermore, for $r < r_c$, because $\zeta(r) = 0$ – the quarks are massless. Except for the nucleon mass, the interaction in the quark-scalar sector does not affect the other predictions of nucleon properties by $NL\sigma M$, which have been quite successful.[31] In particular, the description of baryonic charge distribution as topological charge distribution probed by $\omega$ remains valid. We are therefore led to the physical picture of the nucleon given in Fig. 5. Namely, the nucleon has an outer cloud of $q\bar{q}$ condensed ground state similar to the BCS ground state in a superconductor, an inner core of topological baryonic charge probed by $\omega$, and a still smaller quark-bag where massless valence quarks reside.

At the end of the next section, Sec. 4, we argue that the nucleon structure (Fig. 5) which has emerged implies large $|t|$ $pp$ elastic scattering as due to valence quark-quark scattering. The $pp$ elastic amplitude due to this process is derived in Sec. 5 and the corresponding quantitative calculations are carried out in Sec. 6.

## 4. Asymptotic Description

The developments of Sec. 2 show that to predict elastic differential cross section at some higher energy, such as at LHC, we need to know the energy dependence of $g(s), \Gamma_D^-(s,0)$, and $\exp[i\hat{\chi}(s)]$. Therefore, determining the energy dependence of these parameters is essential for further progress of the model. Against this backdrop, we noticed that model independent analyses of Kundrat and Lokajíček[37,38] showed that at $b = 0$, $\left|\exp[i\chi_D^+(s,0)\right|$ is small but finite and decreases slowly with increasing energy. Specifically, they obtained

$$\tfrac{1}{4}(1-\eta^2(s)) \quad = 0.23 \text{ at } \sqrt{s} = 53 \text{ GeV}$$
$$= 0.24 \text{ at } \sqrt{s} = 541 \text{ GeV},$$

i.e., a small decrease in $\eta(s,0) = \left|\exp[i\chi_D^+(s,0)\right|$ even though the c.m. energy increased by an order of magnitude. This led us to consider the following simple crossing even parametrization [a] for $\exp[i\chi_D^+(s,0)]$:

$$e^{i\chi_D^+(s,0)} = \eta_0 + \frac{c_0}{(se^{-i\pi/2})^\sigma}. \tag{4.1}$$

As $g(s)$ is related to $\exp[i\chi_D^+(s,0)] = 1 - \Gamma_D^+(s,0)$ via Eq. (2.9), the three energy-independent parameters in (4.1) together with the diffraction parameters $R_0, R_1, a_0, a_1$ allow us to obtain $g(s)$ at different values of s:

$$g(s) = \left(1 - \eta_0 - \frac{c_0}{(se^{-i\pi/2})^\sigma}\right)\frac{1+e^{-R/a}}{1-e^{-R/a}}, \tag{4.2}$$

where $R = R(s) = R_0 + R_1(\ln s - i\frac{\pi}{2})$ (Eq. 2.8a) and $a = a(s) = a_0 + a_1(\ln s - i\frac{\pi}{2})$ (Eq. 2.8b).

---

[a] In a preliminary analysis, a somewhat different parametrization was used: M.M. Islam and E.M. Kubik, Proceedings of the VIII[th] Blois Workshop, edited by V.A. Petrov and A. V. Prokudin (World Scientific, 2000) p.325.

The parametrization in Eq. (4.1) indicates that we are in the realm of asymptotic and approach to asymptotic behavior for $\exp[i\chi_D^+(s,0)]$. This, in turn, suggests that we should consider parallel asymptotic and approach to asymptotic behavior for $\Gamma_D^-(s,0)$ and $\exp[i\hat{\chi}(s,0)]$ – all at zero impact parameter. Since $\Gamma_D^-(s,0)$ is crossing odd, we consider for it the following parametrization:

$$\Gamma_D^-(s,0) = i\lambda - i\frac{d_0}{(se^{-i\pi/2})^{\alpha}}\,. \qquad (4.3)$$

On the other hand, $\exp[i\hat{\chi}(s,0)]$ is crossing even as it originates from cloud-cloud diffraction scattering in Fig. 2b, and we take the following parametrization for it:

$$\tilde{\gamma}e^{i\hat{\chi}(s,0)} = \hat{\gamma}_0 + \frac{\hat{\gamma}_1}{(se^{-i\pi/2})^{\hat{\sigma}}}\,. \qquad (4.4)$$

There are, therefore, nine energy-independent parameters now replacing the six energy-dependent parameters of Sec. 2. (We note: $\tilde{\gamma}\exp[i\hat{\chi}(s,0)] = \hat{\gamma}(s)\exp[i\hat{\theta}(s)]$ in Sec. 2.)

The parametrizations (4.1), (4.3), and (4.4) together with those of $R(s)$ and $a(s)$ in Sec. 2 make it evident that a high energy asymptotic description is emerging from the asymptotic and approach to the asymptotic behavior of the elastic scattering amplitude. It is, therefore, appropriate to ask first what asymptotic description we find for our diffraction amplitude $T_D^+(s,t)$:

1. It leads to $\sigma_{tot}(s) \sim (a_0 + a_1\ln s)^2$, i.e. qualitative saturation of the Froissart-Martin bound.
2. It yields $\rho(s) \simeq \pi a_1/(a_0 + a_1\ln s)$ asymptotically, so that the derivative dispersion relation result[39] $\rho(s) = \pi/\ln s$ is satisfied.
3. It obeys the Auberson-Kinoshita-Martin scaling, i.e. $T_D^+(s,t) \sim is\ln^2 s\, f(|t|\ln^2 s)$ asymptotically.
4. It is crossing even, and therefore yields equal $\bar{p}p$ and $pp$ total and differential cross-sections.

The asymptotic properties 1-3 of the diffraction amplitude $T_D^+(s,t)$ can be seen in the following way: Use (2.9) for the profile function in the Fourier-Bessel transform (2.6), change the variable of integration from $b$ to $\zeta = b/a$, and rotate the line of integration a little to the real axis (because $a$ has a small negative imaginary part). This leads to

$$T_D^+(s,t) = ipWg(s)a^2\int_0^{\infty}\zeta\, d\zeta\, J_0(\zeta qa)\frac{\sinh R/a}{\cosh R/a + \cosh\zeta}\,. \qquad (4.5)$$

We observe that, when $a_0 + a_1\ln s \to \infty$,

$$\frac{R_0 + R_1(\ln s - i\pi/2)}{a_0 + a_1(\ln s - i\pi/2)} = \frac{R_0 - ra_0}{a_0 + a_1(\ln s - i\pi/2)} + r \simeq r\,, \qquad (4.6)$$

where $r \equiv R_1/a_1$ is a real quantity. Using this in the integrand in Eq. (4.5), we obtain the first three properties. Property 4 follows by noticing that $ipW \simeq is/2$ is crossing symmetric and $[\Gamma_D^+(se^{i\pi},b)]^* = \Gamma_D^+(s,b)$, so that $T_D^{\bar{p}p}(s,t) = [T_D^{pp}(se^{i\pi},t)]^* = T_D^{pp}(s,t)$.[8] We also note that the combination $a_0 + a_1\ln s$ with $s$ always expressed in our paper in terms of the scale $s_0 = 1$ GeV$^2$ is actually scale invariant, i.e. if $s$ is expressed in terms of a different scale $s_1$ and not $s_0$, then $a_0 + a_1 \ln s = a_0' + a_1 \ln(s/s_1)$, where $a_0' = a_0 + a_1 \ln(s_1/s_0)$.

For $|t| \neq 0$, $T_D^+(s,t)$ is given by the crossing symmetric form[18]

$$T_D^+(s,t) \simeq i\, s\, g(s)\, a\, \tfrac{1}{2}\{-i\pi(R+i\pi a)\, H_0^{(1)}[q(R+i\pi a)] + i\pi(R-i\pi a)\, H_0^{(2)}[q(R-i\pi a)]\}. \quad (4.7)$$

When $q|R \pm i\pi a| >> 1$, the Hankel functions fall off exponentially, which leads to $|T_D^+(s,t)|/s \sim \exp[-q\pi(a_0 + a_1 \ln s)]$ as $s \to \infty$. Since $d\sigma/dt = 4\pi\, |T(s,t)/s|^2$, we find that for $q$ fixed and $s \to \infty$, the differential cross section due to diffraction vanishes. On the other hand, the total cross section due to diffraction tends to infinity as $\ln^2 s$. Hence, Martin's theorem[40] predicts a zero of $\mathrm{Re}\, T_D^+(s,t)$ in the near forward direction. Indeed, in our calculations we find such zeros indicating further the asymptotic nature of our diffraction amplitude.

There are at this point 13 energy independent parameters in the model: $R_0$, $R_1$, $a_0$, $a_1$, $\eta_0, c_0, \sigma, \lambda_0, d_0, \alpha, \hat{\gamma}_0, \hat{\gamma}_1, \hat{\sigma}$. As the model is now expected to provide a high energy asymptotic description, we determined these parameters by requiring that they satisfactorily describe the high energy asymptotic behavior of $\sigma_{tot}(s)$ and $\rho(s)$ given by dispersion relation calculations of Augier et al.,[41] and the experimental $\bar{p}p$ elastic differential cross section data at $\sqrt{s}$ = 546 GeV of Bozzo et al.[2] We then tested the model[21] by calculating $d\sigma/dt$ at $\sqrt{s}$ = 630 GeV and 1.8 TeV and comparing with the experimental $d\sigma/dt$ data.[3,5,6] An adequate description was obtained for both energies, and therefore provided the ground for us to predict $pp$ elastic $d\sigma/dt$ at LHC at the c.m. energy $\sqrt{s}$ =14 TeV. This development showed that the model now has the ability to predict rather than simply fit the data. We note that the profile function (Eq. (2.7)) with the parameters in Ref. 21 has been used by Frankfurt et al.[42] to discuss the impact parameter distribution of soft inelastic collisions in exclusive diffractive Higgs production at LHC.

One aspect of our model, however, remains unaddressed at this point. In $pp$ elastic scattering, when the momentum transfer is Q, a proton probes the other proton at an impact parameter or transverse distance $b \sim 1/Q$. If $Q$ is sufficiently large, then $b$ can be smaller than $2r_c$, where $r_c$ is the critical distance in Fig. 4. One proton then probes the other in a region where quarks are massless and perturbative and the valence quark-bag of one proton overlaps that of the other. Consequently, high energy large $|t|$ elastic $pp$ scattering in our model originates from valence quark-quark elastic scattering in the perturbative regime. The quark counting rules of perturbative QCD[43,44,45] predict at large s and $|t|$: $d\sigma/dt \sim t^{-10}$. On the other hand, Donnachie and Landshoff pointed out that in the energy range $\sqrt{s}$=27 - 62 GeV and $|t| \geq 3.5$ GeV$^2$, $pp$ elastic $d\sigma/dt$ was approximately energy independent and fell off as $t^{-8}$. This was interpreted by them as due to the independent exchanges of three perturbative gluons.[46,47] Later, Sotiropoulos and Sterman[48] pointed out that the three gluons would reggeize, so that color octet exchanges would be suppressed. Instead, three color-singlet exchanges would take their place. Eventually, as $|t|$ increases, a single color-singlet exchange would dominate and lead to $t^{-10}$ fall-off as predicted by the QCD quark-counting rules of Matveev et al.[43] and Brodsky and Farrar.[44] This, of course, means that in our model we expect a change in the behavior of $d\sigma/dt$ from the exponential Orear fall-off: $d\sigma/dt \sim e^{-2\beta\sqrt{|t|}}$ to a power fall-off: $d\sigma/dt \sim t^{-10}$ at some large $|t|$ – resulting in a change in slope of the differential cross section. Such a change in $d\sigma/dt$ slope was observed by De Kerret et al.[49] at ISR for $\sqrt{s}$=53 GeV and $|t| \gtrsim 6.5$ GeV$^2$. However, because of low statistics, experimentally this change in slope could not be reliably established. We now quantitatively address the question of change in the behavior of $d\sigma/dt$ at large $|t|$ due to valence quark-quark elastic scattering in the perturbative regime.

## 5. Valence quark-quark scattering at large $|t|$ and hard pomeron.

As discussed in our paper,[14] we view large $|t|$ $pp$ elastic scattering as a hard collision of a valence quark from one proton with a valence quark from the other proton (Fig. 6). The collision carries off the whole momentum transfer. This dynamical picture brings two new elements into our investigation: The first one is the quark-quark elastic scattering amplitude at high energy and large momentum transfer, which is in the domain of perturbative QCD. The latter has been the focus of extensive studies following the original work of Balitsky, Fadin, Kuraev, and Lipatov (BFKL).[50] The present status is that the $qq$ elastic scattering occurs via reggeized gluon ladders with rungs of gluons which represent gluon emissions in inelastic processes (BFKL pomeron). It is a crossing-even amplitude which is a cut in the angular momentum plane with a fixed branch point at $\alpha_{BFKL} = 1 + \omega$. The value of $\omega$ in the next-to-leading order (NLO) lies in the range 0.13 - 0.18 as argued by Brodsky et al.[51] We refer to the BFKL pomeron with next to leading order corrections included as the QCD "hard pomeron". In our investigation, we approximate this hard pomeron by a fixed pole and take the $qq$ scattering amplitude in Fig. 6 as

$$\hat{T}(\hat{s}, t) = i\gamma_{qq}\,\hat{s}\,(\hat{s}e^{-i\frac{\pi}{2}})^{\omega}\,\frac{1}{|t| + r_0^{-2}}\,, \tag{5.1}$$

where $\hat{s} = (p+k)^2$, $t = -\vec{q}^{\,2}$. The phase in Eq. (5.1) follows from the requirement that $\hat{T}(\hat{s}, t)$ is a crossing even amplitude. Eq. (5.1) represents the hard pomeron amplitude in our calculations. If we want to describe asymptotic quark-quark scattering, we have to take into account unitarity corrections due to infinite exchanges of this hard pomeron. This can be done by taking $\hat{T}(\hat{s}, t)$ as the Born amplitude in an eikonal formulation,[52] which leads to a black-disk description and requires $\gamma_{qq} > 0$. The radius of the black disk turns out to be $R(\hat{s}) = r_0\omega \ln \hat{s}$. Hence, the parameter $r_0$ in Eq. (5.1) has the physical significance of a length scale that defines the black-disk radius of asymptotic quark-quark scattering.

The second element that enters into our consideration is the momentum wave function or probability amplitude $\varphi(\vec{p})$ of a valence quark to have momentum $\vec{p}$ when the proton has momentum $\vec{P}$ in the c.m. system. The way it enters can be seen from Fig. 6. Initially, we have a state where there is a quark of momentum $\vec{p}$ with probability amplitude $\varphi(\vec{p})$ and another quark of momentum $\vec{k}$ with probability amplitude $\varphi(\vec{k})$. Therefore, the initial state is $|i\rangle = \varphi(\vec{p})|\vec{p}\rangle\,\varphi(\vec{k})|\vec{k}\rangle$. The final state after scattering is a quark of momentum $\vec{p} - \vec{q}$ with probability amplitude $\varphi(\vec{p} - \vec{q})$ and a quark of momentum $\vec{k} + \vec{q}$ with probability amplitude $\varphi(\vec{k}+\vec{q})$; therefore, $|f\rangle = \varphi(\vec{p}-\vec{q})|\vec{p}-\vec{q}\rangle\, \varphi(\vec{k}+\vec{q})|\vec{k}+\vec{q}\rangle$. Hence the $pp$ elastic scattering amplitude due to the process shown in Fig. 6 is

$$T_{qq}(s,-\vec{q}^{\,2}) = \sum_{\vec{p},\vec{k}} \varphi^*(\vec{p}-\vec{q})\varphi(\vec{p})\langle\vec{k}+\vec{q}|\langle\vec{p}-\vec{q}|\hat{T}|\vec{p}\rangle|\vec{k}\rangle\varphi^*(\vec{k}+\vec{q})\varphi(\vec{k})\,. \tag{5.2}$$

Now, $\langle\vec{k}+\vec{q}|\langle\vec{p}-\vec{q}|\hat{T}|\vec{p}\rangle|\vec{k}\rangle$ is the $qq$ elastic amplitude corresponding to energy $\hat{s} = (p+k)^2$ and momentum transfer $t = -\vec{q}^{\,2}$. This is the hard pomeron amplitude discussed in the previous paragraph and given by Eq. (5.1). We therefore obtain from (5.2)[b]

$$T_{qq}(s,-\vec{q}^{\,2}) = \sum_{\vec{p},\vec{k}} \varphi^*(\vec{p}-\vec{q})\varphi(\vec{p})\, i\gamma_{qq} \frac{\hat{s}(\hat{s}e^{-i\frac{\pi}{2}})^{\omega}}{\vec{q}^{\,2}+r_0^{-2}}\varphi^*(\vec{k}+\vec{q})\varphi(\vec{k}) \tag{5.3}$$

To evaluate Eq. (5.3), the momentum wave function $\varphi(\vec{p})$ has to be known. We determine it in the following way. From the nucleon structure shown in Fig. 5, we see that the valence quarks in our model are contained in a small quark-bag. We can represent the quark-bag by introducing a probability density $\rho_0(\vec{r})$ of finding a valence quark inside a nucleon at position $\vec{r}$, and then specifying the density, by taking its form factor to be a dipole:

$$F(\vec{q}) = \int \rho_0(\vec{r})e^{i\vec{q}\cdot\vec{r}}d^3r\,, \tag{5.4a}$$

$$= \frac{1}{\left(1+\frac{\vec{q}^{\,2}}{m_0^2}\right)^2}\,. \tag{5.4b}$$

This leads to $\rho_0(\vec{r}) = (m_0^3/8\pi)e^{-m_0 r}$, and indicates that the valence quarks are confined in a region whose r.m.s. radius is $r_c = \sqrt{12}/m_0$. From $\rho_0(\vec{r})$, the momentum wave function $\varphi_0(\vec{p})$ of a valence quark in the rest frame of a proton can now be found.

If $\psi_0(\vec{r})$ is the wave function of a valence quark in the proton rest frame, then $\rho_0(\vec{r}) = |\psi_0(\vec{r})|^2$, and we obtain

$$\psi_0(\vec{r}) = \sqrt{m_0^3/8\pi}\; e^{\frac{-m_0}{2}r}\,. \tag{5.5}$$

$\psi_0(\vec{r})$, on the other hand, has the following expansion

$$\psi_0(\vec{r}) = \sum_{\vec{p}} \varphi_0(\vec{p})\frac{e^{i\vec{p}\cdot\vec{r}}}{\sqrt{V_0}}\,, \tag{5.6}$$

where $\varphi_0(\vec{p})$ is the momentum wave function, or the probability amplitude of a quark to have momentum $\vec{p}$ in the rest frame of the proton; $V_0$ is the quantization volume in the proton rest frame. From Eqs. (5.5) and (5.6), we find

$$\varphi_0(\vec{p}) = \int \frac{e^{-i\vec{p}\cdot\vec{r}}}{\sqrt{V_0}}\psi_0(\vec{r})d^3r \tag{5.7a}$$

---

[b] The RHS of Eq. (5.3) has to be multiplied by a factor of 9 to take into account that there are three valence quarks in each proton. We absorb this factor into the constant $\gamma_{qq}$. The valence quarks in our field theory model are massless effective color-singlet quarks and not the QCD current quarks.

$$= \left( \frac{2\pi m_0^5}{V_0} \right)^{\frac{1}{2}} \frac{1}{(\frac{m_0^2}{4} + \vec{p}^{\,2})^2} \tag{5.7b}$$

The question we now face is: How do we determine $\varphi(\vec{p})$ knowing $\varphi_0(\vec{p})$? This can be done in the following way. If $\rho(\vec{r})$ is the probability density in the c.m. frame, then due to Lorentz contraction

$$\rho(\vec{b} + \vec{e}_3 z) = \gamma \rho_0(\vec{b} + \vec{e}_3 \gamma z)\,, \tag{5.8}$$

where $\gamma$ is the Lorentz contraction factor: $\gamma = E/M = \sqrt{s}/2M$ , $\vec{r} = \vec{b} + \vec{e}_3 z$, $\vec{e}_3$ is the unit vector in the direction of $\vec{P}$, i.e. the z-axis, and M is the nucleon mass. Writing $\rho(\vec{r}) = |\psi(\vec{r})|^2$, where $\psi(\vec{r})$ is the quark wave function in the c.m. frame, we obtain

$$\begin{aligned} \psi(\vec{b} + \vec{e}_3 z) &= \sqrt{\rho(\vec{b} + \vec{e}_3 z)} \\ &= \sqrt{\gamma \rho_0(\vec{b} + \vec{e}_3 \gamma z)} \\ &= \sqrt{\gamma}\, \psi_0(\vec{b} + \vec{e}_3 \gamma z)\,. \end{aligned} \tag{5.9}$$

Again,

$$\psi(\vec{r}) = \sum_{\vec{p}} \varphi(\vec{p}) \frac{e^{i\vec{p}\cdot\vec{r}}}{\sqrt{V}}\,, \tag{5.10}$$

where $\varphi(\vec{p})$ is the momentum wave function in the c.m. frame, i.e. the moving frame, and $V$ is the quantization volume in this frame ($V = V_0/\gamma$). We then get

$$\begin{aligned} \varphi(\vec{p}) &= \int \frac{e^{-i\vec{p}\cdot\vec{r}}}{\sqrt{V}} \psi(\vec{r}) d^3 r \\ &= \sqrt{\gamma} \int \frac{e^{-i(\vec{p}_\perp\cdot\vec{b} + p_3 z)}}{\sqrt{V}} \psi_0(\vec{b} + \vec{e}_3 \gamma z) d^2 b\, dz \\ &= \int \frac{e^{-i(\vec{p}_\perp\cdot\vec{b} + \frac{p_3 z'}{\gamma})}}{\sqrt{V_0}} \psi_0(\vec{b} + \vec{e}_3 z') d^2 b\, dz'\,, \end{aligned} \tag{5.11}$$

which leads to

$$\varphi(\vec{p}_\perp + \vec{e}_3 p_3) = \varphi_0(\vec{p}_\perp + \vec{e}_3 \frac{p_3}{\gamma})\,. \tag{5.12}$$

It is easy to understand this result. Because of Lorentz contraction of the probability density in the moving frame along the z-axis, the corresponding momentum wave function expands along that axis. Eq. (5.12) yields

$$\varphi(\vec{p}_\perp + \vec{e}_3 p_3) = \left( \frac{2\pi m_0^5}{V_0} \right)^{\frac{1}{2}} \left( \frac{m_0^2}{4} + p_\perp^2 + \frac{p_3^2}{\gamma^2} \right)^{-2}. \tag{5.13}$$

We are now in a position to evaluate Eq. (5.3). Introducing light-cone variables $P_+ = P_0 + P_3$, $P_- = P_0 - P_3$, $p_+ = p_0 + p_3$, $p_- = p_0 - p_3$, etc., and writing $p_+ = xP_+$, $k_- = x'K_-$, we find $\hat{s} \simeq xx's$, when $P_+, K_- \to \infty$. Eq. (5.3) then takes the factorizable form

$$T_{qq}(s,-\vec{q}^{\,2}) = \left(\sum_{\vec{p}} \varphi^*(\vec{p}-\vec{q})\varphi(\vec{p})x^{1+\omega}\right) i\gamma_{qq} \frac{s(se^{-i\frac{\pi}{2}})^{\omega}}{\vec{q}^{\,2} + r_0^{-2}} \left(\sum_{\vec{k}} \varphi^*(\vec{k}+\vec{q})\varphi(\vec{k})x'^{1+\omega}\right). \tag{5.14}$$

As shown in appendix B, in the frame where $P_+ \to \infty$,

$$\sum_{\vec{p}} \varphi^*(\vec{p}-\vec{q})\varphi(\vec{p})x^{1+\omega} = \frac{Mm_0^5}{8\pi}\int_0^1 dx \frac{x^{1+\omega}}{(\frac{m_0^2}{4} + M^2x^2)} I(q_\perp, \alpha(x)), \tag{5.15}$$

where

$$I(q,\alpha(x)) = \int_0^\infty b\,db\,J_0(bq)\{bK_1(b\alpha)\}^2 . \tag{5.16}$$

Detailed derivation of Eq. (5.15) and the explicit analytic form of $I(q,\alpha(x))$ and its behavior for small $a'^2 = q^2 / 4\alpha^2(x)$ and $a'^2$ asymptotic are given in appendix B.

Denoting by $\mathscr{F}(q_\perp)$ the right hand side (RHS) of Eq. (5.15), we find $T_{qq}(s,t)$ given by Eq. (5.14) takes the form

$$T_{qq}(s,t) = i\gamma_{qq} s \frac{(se^{-i\frac{\pi}{2}})^{\omega}}{|t| + r_0^{-2}} \mathscr{F}^2(q_\perp), \quad (|t| \simeq \vec{q}_\perp^{\,2}) . \tag{5.17}$$

This is the $pp$ elastic scattering amplitude due to valence quark-quark scattering (Fig. 6). $\mathscr{F}(q_\perp)$ here is not a form factor, even though it resembles one. If $x$ were identically equal to 1 in Eq. (5.15), then it would have been the form factor:

$$\begin{aligned}\sum_{\vec{p}} \varphi^*(\vec{p}-\vec{q})\varphi(\vec{p}) &= F(\vec{q}_\perp + \vec{e}_3 \frac{q_3}{\gamma}) \\ &\simeq F(\vec{q}_\perp)\end{aligned} \tag{5.18}$$

From now on, we refer to $\mathscr{F}(q_\perp)$ as a structure factor to distinguish it from the usual form factor $F(\vec{q}_\perp)$ given by Eq. (5.4).

It is instructive to study the large momentum transfer behavior of $\mathscr{F}(q_\perp)$ and $T_{qq}(s,t)$. In appendix B, we find that for $a'^2 = \frac{q_\perp^2}{4\alpha^2} >> 1$ and $a^2 \simeq a'^2$, Eq. (5.16) yields $I(q_\perp, \alpha(x)) \simeq \frac{4}{q_\perp^4} \simeq \frac{4}{|t|^2}$. Substituting this on the RHS of (5.15), we obtain $\mathscr{F}(q_\perp) \sim \frac{1}{|t|^2}$. Eq. (5.17) then leads to an amplitude

$$T_{qq}(s,t) \sim \frac{i\gamma_{qq}s(se^{-i\frac{\pi}{2}})^{\omega}}{|t|^5} . \tag{5.19}$$

This results in a differential cross section behavior for fixed $s$ and large $|t|$:

$$\frac{d\sigma}{dt} \sim \frac{1}{|t|^{10}} \qquad (s >> |t| >> m_0^2 + 4M^2) . \tag{5.20}$$

Eq. (5.20) shows that we obtain the behavior predicted by the perturbative QCD quark counting rules[43,44,45] for large $|t|$.

## 6. Quantitative Calculations and LHC Results

In our *pp* elastic scattering model, we now have two hard-collision amplitudes: one due to $\omega$ exchange, the other due to hard pomeron exchange. Both collisions are accompanied by cloud-cloud diffraction scattering that reduces these amplitudes by an absorption factor $\exp[i\hat{\chi}(s,0)]$.[18] So the sum of the two hard amplitudes becomes

$$T_1(s,t) = e^{i\hat{\chi}(s,0)}\left[\pm\tilde{\gamma}\, s\frac{F^2(t)}{m_\omega^2 - t} + i\gamma_{qq}\, s\,(s\,e^{-i\pi/2})^{\omega}\,\frac{\mathscr{F}^2(q_\perp)}{|t| + r_0^{-2}}\right], \quad (+\text{ for } \bar{p}p,\ -\text{ for } pp). \quad (6.1)$$

Using the same parameterization as before,[21]

$$\tilde{\gamma}\, e^{i\hat{\chi}(s,0)} = \hat{\gamma}_0 + \frac{\hat{\gamma}_1}{(se^{-i\frac{\pi}{2}})^{\hat{\sigma}}}, \quad (6.2)$$

we find

$$T_1(s,t) = \left[\hat{\gamma}_0 + \frac{\hat{\gamma}_1}{(se^{-i\frac{\pi}{2}})^{\hat{\sigma}}}\right]\left[\pm s\frac{F^2(t)}{m_\omega^2 - t} + i\tilde{\gamma}_{qq}\, s\,(s\,e^{-i\pi/2})^{\omega}\,\frac{\mathscr{F}^2(q_\perp)}{|t| + r_0^{-2}}\right], \quad (6.3)$$

where $\tilde{\gamma}_{qq} = \gamma_{qq}/\tilde{\gamma}$. The $qq$ hard scattering term brings four new parameters: i) $\tilde{\gamma}_{qq}$ which measures the relative strength of this term compared to the $\omega$ exchange term; ii) $\alpha_{BFKL} = 1+\omega$ which controls the high energy behavior; iii) $r_0$ which provides the length scale for the black-disk radius of $qq$ asymptotic scattering; iv) $m_0$ which determines the quark wave function $\psi_0(\vec{r}) = \sqrt{\rho_0(\vec{r})}$ and the size of the quark bag. Because of the different physical aspects associated with them, these four parameters form a minimal set.

We determine the parameters of the model by requiring that the model should describe satisfactorily the asymptotic behavior of $\sigma_{tot}(s)$ and $\rho(s)$ as well as the measured $\bar{p}p$ elastic $d\sigma/dt$ at $\sqrt{s}$ = 546 GeV,[2] 630 GeV,[3] and 1.8 TeV.[5,6] The results of this investigation are shown in Figs. 7 - 9 together with the experimental data. We obtain quite satisfactory descriptions. The dotted curves in Figs. 7 and 8 represent the error bands given by Cudell et. al. (COMPETE Collaboration) to their best fit.[53] We notice that our $\sigma_{tot}(s)$ curve lies within their error band almost overlapping their lower curve, but our $\rho_{pp}(s)$ curve significantly deviates from the error band. Such a deviation is not surprising – since a hard pomeron occurs in our calculations, and the hard pomeron in conjunction with a crossing-odd absorptive correction[18] leads to a crossing-odd amplitude (an odderon).[c] The odderon produces a visible difference between $\rho_{\bar{p}p}(s)$ and $\rho_{pp}(s)$ at large $\sqrt{s}$. The parameters obtained by us for the soft (small $|t|$) diffraction amplitude and the hard (large $|t|$) $\omega$-exchange amplitude are as follows: $R_0 = 2.77$, $R_1 = 0.0491$, $a_0 = 0.245$, $a_1 = 0.126$, $\eta_0 = 0.0844$, $c_0 = 0.00$, $\sigma = 2.70$, $\lambda_0 = 0.727$, $d_0 = 13.0$, $\alpha = 0.246$, $\hat{\gamma}_0 = 1.53$, $\hat{\gamma}_1 = 0.00$, $\hat{\sigma} = 1.46$ (the unit of energy is 1 GeV). The parameters $\beta$ and $m_\omega$ are kept fixed at values found before[16]: $\beta = 3.075$, $m_\omega = 0.801$. There are now seventeen adjustable parameters. The four new parameters describing the hard (large $|t|$) $qq$ amplitude have the values $\tilde{\gamma}_{qq} = 0.03$, $\omega = 0.15$, $r_0 = 2.00$, $m_0^2 = 12.0$. These parameters cannot be determined reliably, because no large $|t|$ elastic data are available in the TeV energy region. The value of $m_0^2$ leads to a valence quark-bag of r.m.s. radius 0.2 F, while that of the baryonic charge core determined from the $\omega$ form factor is 0.44 F.

[c] There is another (albeit small) crossing-odd amplitude in our calculations originating from the crossing-odd $\omega$ exchange amplitude (Eq. (6.1)) multiplied by the crossing-even absorptive correction $[1-\Gamma_D^+(s,0)]$.

Our prediction for $pp$ elastic differential cross section at LHC at $\sqrt{s}$ = 14 TeV for the whole momentum transfer range $|t|$ = 0 – 10 GeV$^2$ is now given in Fig. 10 (solid curve). We obtain for $\sigma_{tot}$ and $\rho_{pp}$ the values 110 mb and 0.120, respectively. Also given in Fig. 10 are separate $d\sigma/dt$ due to diffraction (dotted curve), due to hard $\omega$-exchange (dot-dashed curve), and due to hard $qq$ scattering (dashed curve). As expected in our model, we find that in the small $|t|$ region ($|t| \simeq (0 - 0.5$ GeV$^2$) diffraction dominates, in the intermediate $|t|$ region ($|t| \simeq 1.0 - 4.0$ GeV$^2$) $\omega$-exchange dominates, and in the large $|t|$ region ($|t| \gtrsim 6.0$ GeV$^2$) $qq$ scattering dominates. The three $|t|$ regions correspond to cloud-cloud interaction, core-core scattering due to $\omega$-exchange, and valence $qq$ scattering via QCD hard pomeron. Therefore, they reflect the composite structure of the nucleon with an outer cloud, an inner core of topological baryonic charge, and a still smaller quark-bag of valence quarks.

We mentioned earlier that besides us two other groups, Bourrely et al.[11] and Desgrolard et al.,[12] have predicted $pp$ elastic $d\sigma/dt$ at LHC from $|t| = 0$ to $|t| = 10$ GeV$^2$. In Fig. 11, a comparison of our prediction with those of these authors is given. We see that while our $d\sigma/dt$ (solid curve) falls smoothly after the first dip and second maximum or shoulder, these authors predict identifiable oscillations in $d\sigma/dt$. Furthermore, we predict for $|t| \gtrsim 6$ GeV$^2$ a much larger cross section than them. In Figs. 12 and 13, we show our calculated $pp$ $d\sigma/dt$ at four different energies: $\sqrt{s}$ = 0.5, 1.8, 8 and 14 TeV in the $|t|$ ranges 0 – 10 GeV$^2$ and 0 – 2 GeV$^2$, respectively. These curves show that as the energy increases, $d\sigma/dt$ at $|t|$ = 0 increases, the diffraction peak shrinks, the second maximum or shoulder moves toward smaller $|t|$, and large-$|t|$ $d\sigma/dt$ increases significantly because of the hard pomeron contribution. In Fig. 14, $|e^{i\chi_D^+(s,b)}|$ is shown as a function of $b$ for these energies. We notice that the unitarity limit $|e^{i\chi_D^+(s,b)}| \leq 1$ is always satisfied, the transverse size of the interaction region increases with energy, and even at $b = 0$, absorption is not 100% ($|e^{i\chi_D^+(s,0)}| > 0$). We also find that $\sigma_{el}(s)/\sigma_{tot}(s)$ tends to 0.2 for $\sqrt{s}$ $\gtrsim 10$ TeV. This, of course, means that there is no sign of black disk behavior, which predicts $\sigma_{el}(s)/\sigma_{tot}(s) \to 0.5$. In Fig. 15, we plot the phases of the diffraction amplitude $T_D^+(s,t)$ (dashed curve) and the full amplitude $T(s,t)$ (solid curve) as functions of $|t|$ for LHC energy: $\sqrt{s}$ = 14 TeV. We find that near the origin the phase of $T_D^+(s,t)$ increases through $\pi/2$ indicating a zero of Re $T_D^+(s,t)$ as expected from Martin's theorem[40] and discussed in Sec. 4. The position of this zero is at $|t| = 0.0781$ GeV$^2$. Re $T(s,t)$ also has a zero nearby at $|t| = 0.0652$ GeV$^2$. Finally, Figs. 16 and 17 show recent $d\sigma/dt$ calculations by Kaspar[54] at $\sqrt{s}$ = 541 GeV ( $\bar{p}p$ ) and 14 TeV ( $pp$ ) in the very small $|t|$ region with the Coulomb interaction included in our model. Kaspar obtains the two rising curves in each figure using for the combined Coulomb-hadronic amplitude the West-Yennie formulation (lower curve) and the Kundrat- Lokajíček formulation (upper curve).[37] The straight line in each figure represents $d\sigma/dt$ due to our hadronic amplitude alone. The experimental data in Fig. 16 are from Augier et al.[55]

## 7. Concluding Remarks

Our phenomenological study of $pp$ and $\bar{p}p$ elastic scattering during the last few years has led to two important developments of the earlier work. First, emergence of an asymptotic description that incorporates the asymptotic and the approach to the asymptotic behavior of the elastic scattering amplitude. This has allowed us to predict $pp$ elastic scattering at LHC at 14 TeV c.m. energy – once the energy independent parameters of the model are determined from the known asymptotic behavior of $\sigma_{tot}(s)$ and $\rho(s)$, and the experimentally measured $\bar{p}p$ $d\sigma/dt$ at $\sqrt{s}$ = 546 GeV, 630 GeV, and 1.8 TeV. Second, description of large $|t|$ elastic scattering in terms of valence quark-quark scattering where the valence quarks are confined in a small region within the proton and scatter via hard pomeron (BFKL pomeron plus next to leading order corrections). Because of the hard pomeron, the large $|t|$ differential cross section increases rapidly with energy. We envision the possibility that this energy dependence can be tested, if the LHC is run at $\sqrt{s}$ =1.8 and 8 TeV besides 14 TeV.

Comparison of our predicted elastic differential cross section at LHC c.m. energy 14 TeV based on the nucleon-structure model with the predictions of the impact-picture model and the eikonalized pomeron-reggeon model (Fig. 11) shows that for $|t| > 1$ GeV$^2$, we predict a smooth fall of $d\sigma/dt$ all the way to $|t| = 10$ GeV$^2$, whereas the other two models predict noticeable diffraction oscillations. Furthermore, we predict for $|t| \gtrsim 6\,\text{GeV}^2$, a much larger cross section than the other two models. Precise measurement of $pp$ elastic $d\sigma/dt$ from $|t| = 0$ to $|t| \simeq 10$ GeV$^2$ by the TOTEM group will be able to distinguish between these different predictions. This, of course, presents a challenging experimental task given that $d\sigma/dt$ falls off by ten orders of magnitude from $|t| = 0$ to $|t| = 10$ GeV$^2$.

Our recent phenomenological studies together with the earlier ones, now extending over twenty-five years, have led to the following physical picture of the nucleon (Fig. 5): The nucleon is a composite object with an outer cloud of quark-antiquark ground state analogous to a superconducting ground state, an inner core of topological baryonic charge (size $\simeq$ 0.44F) which is probed by $\omega$, and a still smaller quark-bag of valence quarks (size $\simeq$ 0.2 F). We have found that a gauged Gell-Mann-Levy type linear $\sigma$-model emerges as an effective field theory model underlying this nucleon structure. The model extends the usual soliton description of the nonlinear $\sigma$-model with Wess-Zumino-Witten action by bringing in a quark-scalar sector where quarks and antiquarks interact via a scalar field. The quantum fluctuation of this scalar field can be identified with the $\sigma$-field of single-boson exchange potentials ($m_\sigma \simeq 500$ MeV), which provides the medium range $NN$ interaction.[56] We note that a scalar field has also been introduced in the context of QCD trace anomaly – when chiral symmetry is spontaneously broken in the presence of light quarks.[57]

Various physical aspects of the emerging nucleon structure have been proposed and studied by other authors in different contexts. For example, the idea of quark-antiquark interaction leading to a $q\bar{q}$ ground state analogous to a BCS superconducting ground state goes back to Nambu-Jona-Lasinio.[58,59] The idea of baryonic charge being topological was proposed by Skyrme,[60] which led to the extensive study of the nucleon as a topological soliton.[31,61] The possibility that baryonic charge can be partly topological and partly due to valence quarks confined in a spherical bag gave birth to the chiral bag model of the nucleon.[62,63] What our investigation shows is that this chiral bag lies embedded in a $q\bar{q}$ condensed ground state and the latter forms the outer cloud of the nucleon. The birth of the quark model in the mid-sixties signaled that the nucleon is a composite object. Since then over the last four decades an enormous amount of effort has been devoted by physicists to determine this composite structure. If LHC corroborates the nucleon structure we find, then it will establish that many of the novel ideas of that effort blend together to provide a unique description of the nucleon.

**Acknowledgment**

One of us (MMI) has greatly benefited from stimulating discussions with many colleagues over the years. He particularly wishes to thank Ngee-Pong Chang, Jerry Friedman, Maurice Jacob, Vojtech Kundrat, George Sterman, and R. Vinh Mau. RJL wishes to thank Ben Luddy and Rick Luddy for their help with programming for data acquisition and artwork. We thank Jan Kaspar for communicating his results to us.

**Appendix A**

To derive the anomalous action of the action functional or effective action $W[A^L(x)]$, we start with its path integral representation in Euclidean space:

$$e^{iW[A^L(x)]} = \frac{1}{\mathcal{N}_L}\int d\psi_L\, d\overline{\psi}_L\; e^{\int d^4x \overline{\psi}_L(x)\gamma^\mu(\partial_\mu + A^L_\mu(x))\psi_L(x)} . \tag{A.1}$$

$\mathcal{N}_L$ is a normalization constant chosen such that the free field term is subtracted out. Next we consider extending the gauge field $A^L_\mu(x)$ to a fifth dimension:

$$\mathscr{A}^L_\mu(x,t) = U_L(x,t)\,A^R_\mu(x)U^\dagger_L(x,t) + U_L(x,t)\,\partial_\mu U^\dagger_L(x,t)\,, \tag{A.2}$$

where $t$ is a parameter that lies between 0 and 1. $U_L(x,t)$ is a unitary matrix, and we require $U_L(x,t)$ to satisfy the conditions

$$U_L(x,t) = \xi^\dagger_L(x)\,\xi_R(x)\,, \qquad (t=1) \tag{A.3a}$$

$$= \xi_R(x)\,. \qquad (t=0) \tag{A.3b}$$

Since $A^R_\mu(x) = \xi^\dagger_R(x)\mathcal{V}_\mu(x)\xi_R(x) + \xi^\dagger_R(x)\partial_\mu\xi_R(x)$, (Eq. (3.16)), the conditions (A.3a,b) lead to

$$\begin{aligned}\mathscr{A}^L_\mu(x,1) &= \xi^\dagger_L(x)\,\xi_R(x)\,A^R_\mu(x)\,\xi^\dagger_R(x)\,\xi_L(x) + \xi^\dagger_L(x)\,\xi_R(x)\,\partial_\mu(\xi^\dagger_R(x)\,\xi_L(x))\\ &= \xi^\dagger_L(x)\mathcal{V}_\mu(x)\,\xi_L(x) + \xi^\dagger_L(x)\,\partial_\mu\xi_L(x)\\ &= A^L_\mu(x)\,,\end{aligned} \tag{A.4}$$

and

$$\begin{aligned}\mathscr{A}^L_\mu(x,0) &= \xi_R(x)\,A^R_\mu(x)\,\xi^\dagger_R(x) + \xi_R(x)\,\partial_\mu\xi^\dagger_R(x)\\ &= \mathcal{V}_\mu(x)\,.\end{aligned} \tag{A.5}$$

Therefore, the matrix $U_L(x,t)$ interpolates the field $\mathscr{A}^L_\mu(x,t)$ between $\mathcal{V}_\mu(x)$ and $A^L_\mu(x)$ as $t$ varies from 0 to 1.

Next we consider the parameter dependent effective action:

$$e^{iW[\mathscr{A}^L(x,t)]} = \frac{1}{\mathcal{N}_L}\int d\psi_L\, d\overline{\psi}_L\, e^{\int d^4x \overline{\psi}_L\gamma^\mu(\partial_\mu + \mathscr{A}^L_\mu(x,t))\psi_L(x)} \tag{A.6a}$$

$$= \frac{1}{\mathcal{N}_L}\int d\psi_L\, d\overline{\psi}_L\, e^{\int d^4x \overline{\psi}_L\gamma^\mu U_L(x,t)(\partial_\mu + A^R_\mu(x))U^\dagger_L(x,t)\psi_L(x)} . \tag{A.6b}$$

From (A.6b),

$$\begin{aligned}e^{iW[\mathscr{A}^L(x,t+\delta t)]} &= \frac{1}{\mathcal{N}_L}\int d\psi_L\, d\overline{\psi}_L\, e^{\int d^4x \overline{\psi}_L\gamma^\mu U_L(x,t+\delta t)(\partial_\mu + A^R_\mu(x))U^\dagger_L(x,t+\delta t)\psi_L(x)}\\ &= \frac{1}{\mathcal{N}_L}\int d\psi_L\, d\overline{\psi}_L\, e^{\int d^4x \overline{\psi}_L(1+\partial_t U_L U^\dagger_L\delta t)\gamma^\mu\,(\partial_\mu + \mathscr{A}^L_\mu(x,t))(1+U_L\,\partial_t\,U^\dagger_L\delta t)\psi_L(x)}\\ &= \frac{1}{\mathcal{N}_L}\int d\psi_L\, d\overline{\psi}_L\, e^{\int d^4x \overline{\psi}'_L\gamma^\mu(\partial_\mu + \mathscr{A}^L_\mu(x,t))\,\psi'_L(x)}\,,\end{aligned} \tag{A.7}$$

where

$$\psi'_L(x) = (1 + U_L\,\partial_t U^\dagger_L\,\delta t)\,\psi_L(x)\,, \tag{A.8a}$$

$$\overline{\psi}'_L(x) = \overline{\psi}_L(x)(1 + \partial_t U_L U^\dagger_L\,\delta t)\,. \tag{A.8b}$$

We now change Grassmann integration variables from $\psi_L$ and $\overline{\psi}_L$ to $\psi'_L$ and $\overline{\psi}'_L$. The Jacobian of this transformation can be calculated following Fujikawa's approach.[64,65] The details are given in ref. 36. The result is

$$d\psi_L\, d\overline{\psi}_L = e^{-\int d^4x\, tr[U_L^\dagger \partial_t U_L \gamma^5 \sum_n \varphi_n(x,t)\varphi_n^\dagger(x,t)]\delta t}\, d\psi'_L\, d\overline{\psi}'_L, \tag{A.9}$$

where $\varphi_n(x,t)$'s form a complete set of eigenfunctions of the hermitian operator $\gamma^\mu(\partial_\mu + \mathscr{A}_\mu^L(x,t))$:

$$\gamma^\mu(\partial_\mu + \mathscr{A}_\mu^L(x,t))\,\varphi_n(x,t) = \lambda_n(t)\,\varphi_n(x,t)\,. \tag{A.10}$$

From (A.7) and (A.9), we obtain

$$e^{iW[\mathscr{A}^L(x,t+\delta t)]} = e^{-\int d^4x\, tr[U_L^\dagger \partial_t U_L \gamma^5 \sum_n \varphi_n(x,t)\varphi_n^\dagger(x,t)]\delta t}\, e^{iW[\mathscr{A}_L(x,t)]}\,. \tag{A.11}$$

Therefore, we find

$$i\,\frac{\partial W[\mathscr{A}^L(x,t)]}{\partial t} = -\int d^4x\;\; tr_G\,[U_L^\dagger(x,t)\;\partial_t U_L(x,t)\;\; tr_D\,(\gamma^5 \sum_n \varphi_n(x,t)\varphi_n^\dagger(x,t)\,)\,], \tag{A.12}$$

where $tr_G$ and $tr_D$ are traces over the group and Dirac matrices.

Again, by directly differentiating (A.6a) with respect to $t$,

$$i\,\frac{\partial W[\mathscr{A}^L(x,t)]}{\partial t} = \int d^4x \left\langle \overline{\psi}_L\, \gamma^\mu\,(-iT^a)\psi_L \right\rangle \dot{\mathscr{A}}_\mu^{L,a}(x,t)$$

$$= \int d^4x\, j_L^{\mu,a}(x,t)\,\dot{\mathscr{A}}_\mu^{L,a}(x,t)\,, \tag{A.13}$$

where $j_L^{\mu,a}(x,t)$ is the left current

$$j_L^{\mu,a}(x,t) = i\,\frac{\delta W[\mathscr{A}^L(x,t)]}{\delta \mathscr{A}_\mu^{L,a}(x,t)} \tag{A.14a}$$

$$= \left\langle \overline{\psi}_L\;\gamma^\mu\,(-iT^a)\psi_L \right\rangle. \tag{A.14b}$$

Eq. (A.13) can be written as

$$i\,\frac{\partial W[\mathscr{A}^L(x,t)]}{\partial t} = -2\int d^4x\; tr_G\,[\,j_L^\mu(x,t)\dot{\mathscr{A}}_\mu^L(x,t)]\,; \tag{A.15}$$

$$j_L^\mu(x,t) = -iT^a\, j_L^{\mu,a}(x,t), \qquad \mathscr{A}_\mu^L(x,t) = -iT^a\, \mathscr{A}_\mu^{L,a}(x,t)\,.$$

Now, from (A.2),

$$\dot{\mathscr{A}}_\mu^L(x,t) = \partial_t \mathscr{A}_\mu^L(x,t)$$

$$= D_\mu(\mathscr{A}^L)\,(U_L^\dagger(x,t)\partial_t U_L(x,t))\,, \tag{A.16}$$

where $D_\mu(\mathscr{A}^L)$ is the covariant derivative: $D_\mu(\mathscr{A}^L) = \partial_\mu + [\mathscr{A}_\mu^L,\quad]$. Inserting (A.16) in (A.15) and carrying out a partial integration, we obtain

$$i\,\frac{\partial W[\mathscr{A}^L(x,t)]}{\partial t} = 2\int d^4x\; tr_G\,[U_L^\dagger(x,t)\partial_t U_L(x,t)\; D_\mu(\mathscr{A}^L)\; j_L^\mu(x,t)]\,. \tag{A.17}$$

Comparing (A.17) with (A.12), we find

$$D_\mu(\mathscr{A}^L)\; j_L^\mu(x,t) = -\tfrac{1}{2} tr_D\,(\gamma^5 \sum_n \varphi_n(x,t)\varphi_n^\dagger(x,t))\,. \tag{A.18}$$

The RHS of (A.18) is not well defined and has to be regularized. Denoting the regularized expression by $a_L(\mathscr{A}^L)$, we get[36]

$$D_\mu(\mathscr{A}^L)\, j_L^\mu(x,t) = a_L(\mathscr{A}^L)\,, \tag{A.19}$$

where

$$a_L(\mathscr{A}^L) = \frac{1}{48\pi^2}\epsilon^{\lambda\mu\rho\sigma}\,\partial_\lambda\,(\mathscr{A}_\mu^L\,\partial_\rho\,\mathscr{A}_\sigma^L + \tfrac{1}{2}\mathscr{A}_\mu^L\,\mathscr{A}_\rho^L\,\mathscr{A}_\sigma^L)\,. \tag{A.20}$$

(A.20) is the Bardeen anomaly for the left-handed current in a model where the action is the same as ours.[66] Integrating (A.17), we finally obtain

$$e^{iW[A^L(x)]} = e^{2\int_0^1 dt\int d^4x\, tr_G[U_L^\dagger(x,t)\partial_t U_L(x,t)\, a_L(\mathscr{A}^L)]}\; e^{iW_L[\mathcal{V}(x)]}\,, \tag{A.21}$$

where

$$e^{iW_L[\mathcal{V}(x)]} = \frac{1}{\mathcal{N}_L}\int d\psi_L\, d\overline{\psi}_L\; e^{\int d^4x\,\overline{\psi}_L(x)\gamma^\mu(\partial_\mu+\mathcal{V}_\mu(x))\psi_L(x)}\,. \tag{A.22}$$

The exponent of the first term on the RHS of (A.21) gives the anomalous action of the quantum functional $W[A^L(x)]$.To derive the anomalous action of the functional $W[A^R(x)]$, we proceed in a similar fashion as above. Starting with the path integral

$$e^{iW[A^R(x)]} = \frac{1}{\mathcal{N}_R}\int d\psi_R\, d\overline{\psi}_R\; e^{\int d^4x\,\overline{\psi}_R(x)\gamma^\mu(\partial_\mu+A_\mu^R(x))\psi_R(x)}\,, \tag{A.23}$$

we first extend the gauge field $A_\mu^R(x)$ to a fifth dimension:

$$\mathscr{A}_\mu^R(x,t) = U_R(x,t)\,A_\mu^L(x)U_R^\dagger(x,t) + U_R(x,t)\partial_\mu U_R^\dagger(x,t)\,. \tag{A.24}$$

The unitary matrix $U_R(x,t)$ is required to satisfy the conditions

$$U_R(x,1) = \xi_R^\dagger(x)\,\xi_L(x)\,,\quad (t=1) \tag{A.25a}$$

$$= \xi_L(x)\,,\qquad (t=0)\,. \tag{A.25b}$$

Again, as $A_\mu^L(x) = \xi_L^\dagger(x)\mathcal{V}_\mu(x)\,\xi_L(x) + \xi_L^\dagger(x)\partial_\mu\xi_L(x)$, we find

$$\begin{aligned}\mathscr{A}_\mu^R(x,1) &= \xi_R^\dagger(x)\,\xi_L(x)\,A_\mu^L(x)\,\xi_L^\dagger(x)\,\xi_R(x) + \xi_R^\dagger(x)\,\xi_L(x)\,\partial_\mu\left(\xi_L^\dagger(x)\,\xi_R(x)\right)\\ &= \xi_R^\dagger(x)\mathcal{V}_\mu(x)\,\xi_R(x) + \xi_R^\dagger(x)\partial_\mu\xi_R(x)\\ &= A_\mu^R(x)\end{aligned} \tag{A.26}$$

and

$$\begin{aligned}\mathscr{A}_\mu^R(x,0) &= \xi_L(x)\,A_\mu^L(x)\,\xi_L^\dagger(x) + \xi_L(x)\partial_\mu\xi_L^\dagger(x)\\ &= \mathcal{V}_\mu(x)\,.\end{aligned} \tag{A.27}$$

Therefore, the matrix $U_R(x,t)$ interpolates the gauge field $\mathscr{A}_\mu^R(x,t)$ between $\mathcal{V}_\mu(x)$ and $A_\mu^R(x)$ as $t$ goes from 0 to 1.

The parameter dependent effective action in this case is

$$e^{iW[\mathscr{A}^R(x,t)]} = \frac{1}{\mathcal{N}_R}\int d\psi_R\, d\overline{\psi}_R\; e^{\int d^4x\,\overline{\psi}_R(x)\gamma^\mu(\partial_\mu+\mathscr{A}_\mu^R(x,t))\psi_R(x)} \tag{A.28a}$$

$$= \frac{1}{\mathcal{N}_R}\int d\psi_R\, d\overline{\psi}_R\; e^{\int d^4x\,\overline{\psi}_R(x)U_R(x,t)\gamma^\mu(\partial_\mu+A_\mu^L(x))U_R^\dagger(x,t)\psi_R(x)}\,. \tag{A.28b}$$

As in (A.7), we consider an infinitesimal increase in the parameter $t$, and then change the Grassmann integration variables to

$$\psi'_R(x) = (1 + U_R\, \partial_t U_R^\dagger\, \delta t)\psi_R(x)\,, \tag{A.29a}$$

$$\bar{\psi}'_R(x) = \bar{\psi}_R(x)(1 + \partial_t U_R\, U_R^\dagger\, \delta t)\,. \tag{A.29b}$$

The Jacobian of the transformation can be calculated as before, and we find

$$i\,\frac{\partial W[\mathscr{A}^R(x,t)]}{\partial t} = \int d^4x\ tr_G\,[\,U_R^\dagger(x,t)\ \partial_t U_R(x,t)\ tr_D\,(\gamma^5 \sum_n \tilde{\varphi}_n(x,t)\tilde{\varphi}_n^\dagger(x,t))\,]\,, \tag{A.30}$$

where $\tilde{\varphi}_n(x,t)$'s are the eigenfunctions of the hermitian operator $\gamma^\mu\,(\partial_\mu + \mathscr{A}_\mu^R(x,t))$:

$$\gamma^\mu\,(\partial_\mu + \mathscr{A}_\mu^R(x,t))\tilde{\varphi}_n(x,t) = \tilde{\lambda}_n(t)\,\tilde{\varphi}_n(x,t)\,. \tag{A.31}$$

Carrying out direct differentiation of $iW[\mathscr{A}^R(x,t)]$ with respect to $t$ and then following steps parallel to (A.13) – (A.18), we arrive at the result

$$D_\mu(\mathscr{A}^R)\ j_R^\mu(x,t) = \tfrac{1}{2} tr_D\,(\gamma^5 \sum_n \tilde{\varphi}_n(x,t)\tilde{\varphi}_n^\dagger(x,t))\,. \tag{A.32}$$

An important difference in sign between (A.18) and (A.32) occurs. Denoting the regularized expression for the RHS of (A.32) by $a_R(\mathscr{A}^R)$, we get

$$D_\mu(\mathscr{A}^R)\ j_R^\mu(x,t) = a_R(\mathscr{A}^R)\,, \tag{A.33}$$

where

$$a_R(\mathscr{A}^R) = -\frac{1}{48\pi^2}\epsilon^{\lambda\mu\rho\sigma}\,\partial_\lambda\,(\mathscr{A}_\mu^R\,\partial_\rho\,\mathscr{A}_\sigma^R + \tfrac{1}{2}\mathscr{A}_\mu^R\,\mathscr{A}_\rho^R\,\mathscr{A}_\sigma^R)\,. \tag{A.34}$$

This is the Bardeen anomaly for the right-handed current and has indeed an overall sign opposite to that of the left-handed current.[66] Integrating (A.30), we arrive at

$$e^{iW[A^R(x)]} = e^{2\int_0^1 dt \int d^4x\ tr_G[U_R^\dagger(x,t)\partial_t U_R(x,t)\ a_R(\mathscr{A}^R)]}\ e^{iW_R[\mathcal{V}(x)]}\,, \tag{A.35}$$

and

$$e^{iW_R[\mathcal{V}(x)]} = \frac{1}{\mathcal{N}_R}\int d\psi_R\, d\bar{\psi}_R\, e^{\int d^4x \bar{\psi}_R(x)\gamma^\mu(\partial_\mu + \mathcal{V}_\mu(x))\psi_R(x)}\,. \tag{A.36}$$

The exponent of the first term on the RHS of (A.35) now gives the anomalous part of the action functional $W[A^R(x)]$.

We now consider the possibility of choosing a single interpolating matrix $U(x,t)$ such that $U_L(x,t) = U(x,t)$ and $U_R(x,t) = U^\dagger(x,t)$. Comparing (A.3a,b) and (A.25a,b), we find this requires $\xi_L(x) = \xi_R^\dagger(x)$, i.e., we have to choose the unitary gauge: $\xi_L^\dagger(x) = \xi_R(x) = \xi(x)$, $U(x,1) = \xi^2(x) = U(x)$. Eqs. (A.21) and (A.35) can then be combined, which leads to

$$e^{i(W[A^L(x)] + W[A^R(x)])} = e^{i\Gamma_{WZW}[U,\mathcal{V}]}\ e^{i(W_L[\mathcal{V}(x)] + W_R[\mathcal{V}(x)])}\,, \tag{A.37}$$

where the anomalous Wess-Zumino-Witten action is given by

$$i\Gamma_{WZW}[U,\mathcal{V}] = 2\int_0^1 dt \int d^4x\ tr_G\,[U^\dagger(x,t)\,\partial_t U(x,t)\,a_L(\mathscr{A}^L) + U(x,t)\partial_t U^\dagger(x,t)\,a_R(\mathscr{A}^R)]\,. \tag{A.38}$$

Because the left and the right anomalies have opposite signs (Eqs. (A.20) and (A.34)),

$$i\Gamma_{WZW}[U,\mathcal{V}] \to -i\Gamma_{WZW}[U,\mathcal{V}]\,, \tag{A.39}$$

when

$$U(x,t) \to U^\dagger(x,t)\,,$$
$$\mathscr{A}_\mu^L(x,t) \to \mathscr{A}_\mu^R(x,t)\,,$$
$$\mathscr{A}_\mu^R(x,t) \to \mathscr{A}_\mu^L(x,t)\,.$$

This property of $\Gamma_{WZW}[U,\mathcal{V}]$ in left-right form has been pointed out by Pak and Rossi.[67] Eqs. (A.37) and (A.38) occur in Sec. 3 as Eqs. (3.21) and (3.22).

**Appendix B**

As a first step to derive Eq. (5.15), we consider the following quantity

$$\mathscr{J} = \sum_{\vec{p}} \varphi^*(\vec{p}-\vec{q})\varphi(\vec{p})$$

$$= V\int \frac{d^2 p_\perp}{(2\pi)^2}\int_0^{P_3} \frac{dp_3}{2\pi}\varphi_0^*(\vec{p}_\perp - \vec{q}_\perp + \vec{e}_3\frac{p_3-q_3}{\gamma})\,\varphi_0(\vec{p}_\perp + \vec{e}_3\frac{p_3}{\gamma}) \qquad \text{(using Eq. (5.12))}$$

$$= V\int \frac{d^2 p_\perp}{(2\pi)^2}\int_0^{P_3} \frac{dp_3}{2\pi}\frac{1}{V_0}\left(\frac{m_0^3}{8\pi}\right)\int e^{i(\vec{p}_\perp-\vec{q}_\perp)\cdot\vec{b}'+i\frac{p_3}{\gamma}z'}\, e^{-\frac{m_0}{2}\sqrt{b'^2+z'^2}}\, d^2b'\,d^2z'$$

$$\times\int e^{-i\vec{p}_\perp\cdot\vec{b}-i\frac{p_3}{\gamma}z}\, e^{-\frac{m_0}{2}\sqrt{b^2+z^2}}\, d^2b\,dz \qquad \text{(using Eqs. (5.5) and (5.7a))} \qquad \text{(B.1)}$$

The integral over z, for example, can be easily obtained by noticing that it reduces to $2\int_0^\infty \cos(\frac{p_3 z}{\gamma})e^{-\frac{m_0}{2}\sqrt{b^2+z^2}}dz$, which is a known integral. We find

$$\mathscr{J} = V\int \frac{d^2 p_\perp}{(2\pi)^2}\int_0^{P_3} \frac{dp_3}{2\pi}\frac{1}{V_0}\left(\frac{m_0^3}{8\pi}\right)$$

$$\times 4\int e^{i(\vec{p}_\perp-\vec{q}_\perp)\cdot\vec{b}'}\left(\frac{b'm_0}{2}\right)\frac{K_1[b'(\frac{m_0^2}{4}+\frac{p_3^2}{\gamma^2})^{\frac{1}{2}}]}{(\frac{m_0^2}{4}+\frac{p_3^2}{\gamma^2})^{\frac{1}{2}}}d^2b'$$

$$\times\int e^{-i\vec{p}_\perp\cdot\vec{b}}\left(\frac{bm_0}{2}\right)\frac{K_1[b(\frac{m_0^2}{4}+\frac{p_3^2}{\gamma^2})^{\frac{1}{2}}]}{(\frac{m_0^2}{4}+\frac{p_3^2}{\gamma^2})^{\frac{1}{2}}}d^2b$$

$$= V\frac{m_0^2}{4\pi}P_+\int_0^1 dx\frac{1}{V_0}\left(\frac{m_0^3}{8\pi}\right)\int e^{-i\vec{q}_\perp\cdot\vec{b}}\left\{\frac{bK_1[b\alpha]}{\alpha}\right\}^2 d^2b \qquad \text{(B.2)}$$

using $p_3 \simeq \frac{1}{2}xP_+$, $\frac{p_3}{\gamma}\simeq Mx$ $\quad(\gamma = E/M = \sqrt{M^2+P_3^2}\big/M)$, and writing

$\alpha = \alpha(x) = (\frac{m_0^2}{4}+M^2x^2)^{1/2}$.

Carrying out the angular integration in the b-plane, we finally get

$$\mathscr{J} = \left(\frac{Mm_0^5}{8\pi}\right)\int_0^1 \frac{dx}{(\frac{m_0^2}{4}+M^2x^2)}\, I(q_\perp,\alpha(x))\ , \qquad \text{(B.3)}$$

where

$$I(q,\alpha) = \int_0^\infty b\,db\,J_0(bq)\{bK_1[b\alpha]\}^2\ . \qquad \text{(B.4)}$$

From the definition of $\mathscr{J}$ in (B.1), and (B.3), Eq. (5.15) follows automatically.

An explicit analytic form of $I(q,\alpha)$ as a function of $q$ and $\alpha$ can be derived. Eq. (B.4) can be expressed as

$$I(q,\alpha) = 2\int_0^\infty b^3\,db\,J_0(bq)\int_0^\infty K_2[2b\alpha\cosh t]dt$$

$$= 16\int_0^\infty (2\alpha \cosh t)^2 \frac{dt}{(q^2 + 4\alpha^2 \cosh^2 t)^3}$$

$$= -16\left[\frac{d}{dq^2}\mathscr{I}(q) + \tfrac{1}{2}q^2 \frac{d^2}{d(q^2)^2}\mathscr{I}(q)\right] \quad \text{(B.5)}$$

where
$$\mathscr{I}(q) = \int_0^\infty \frac{dt}{(q^2 + 4\alpha^2 \cosh^2 t)} \quad \text{(B.6)}$$

Carrying out the integration in (B.6), we get

$$\mathscr{I}(q) = \frac{1}{4\alpha^2}\frac{1}{aa'}\ln(a' + a) \quad \text{(B.7)}$$

where $a'^2 = \dfrac{q^2}{4\alpha^2}$, $\quad a^2 = a'^2 + 1$.

After carrying out the derivatives in (B.5), we find

$$I(q,\alpha) = \frac{1}{8\alpha^4}\left\{\left[\frac{2}{a^2} + \frac{1}{a'^2} - \frac{3a'^2}{a^4}\right]\frac{1}{aa'}\ln(a' + a) - \frac{1}{a^2 a'^2} + \frac{3}{a^4}\right\} \quad \text{(B.8)}$$

Determining the behavior of $I(q,\alpha)$ for $q \to 0$, i.e. $a' \to 0$ $\quad (a' = q/2\alpha)$ needs care, since (B.8) tends to suggest $I(q,\alpha)$ to be singular in this limit. The key is to use the following power series expansion[68]

$$\frac{1}{aa'}\ln(a' + a) = \frac{1}{a'\sqrt{a'^2 + 1}}\ln(a' + \sqrt{a'^2 + 1})$$

$$= \sum_{k=0}^\infty (-1)^k \frac{2^{2k}(k!)^2}{(2k+1)!}a'^{2k} \quad \text{(B.9)}$$

A systematic expansion of (B.8) in powers of $a'^2$ then yields for $a' \to 0$,

$$I(q,\alpha) = \frac{1}{8\alpha^4}\left\{\frac{16}{3} - \frac{64a'^2}{5} + \frac{768a'^4}{35} + ...\right\} \quad \text{(B.9a)}$$

$$= \frac{1}{8\alpha^4}\left\{\frac{16}{3} - \frac{16q^2}{5\alpha^2} + \frac{48q^4}{35\alpha^4} + ...\right\}, \quad \text{(B.9b)}$$

which shows that $I(q,\alpha)$ is finite at $q^2 = 0$. The behavior of $I(q,\alpha)$ for $q \to \infty$, i.e. $a' \to \infty$, is much simpler to see: From (B.8),

$$I(q,\alpha) \simeq \frac{1}{8\alpha^4}\left\{\frac{4}{a'^4}\cdot\frac{1}{a'^2}\ln 2a' + \frac{2}{a'^4}\right\}$$

$$\simeq \frac{4}{q^4}. \quad \text{(B.10)}$$

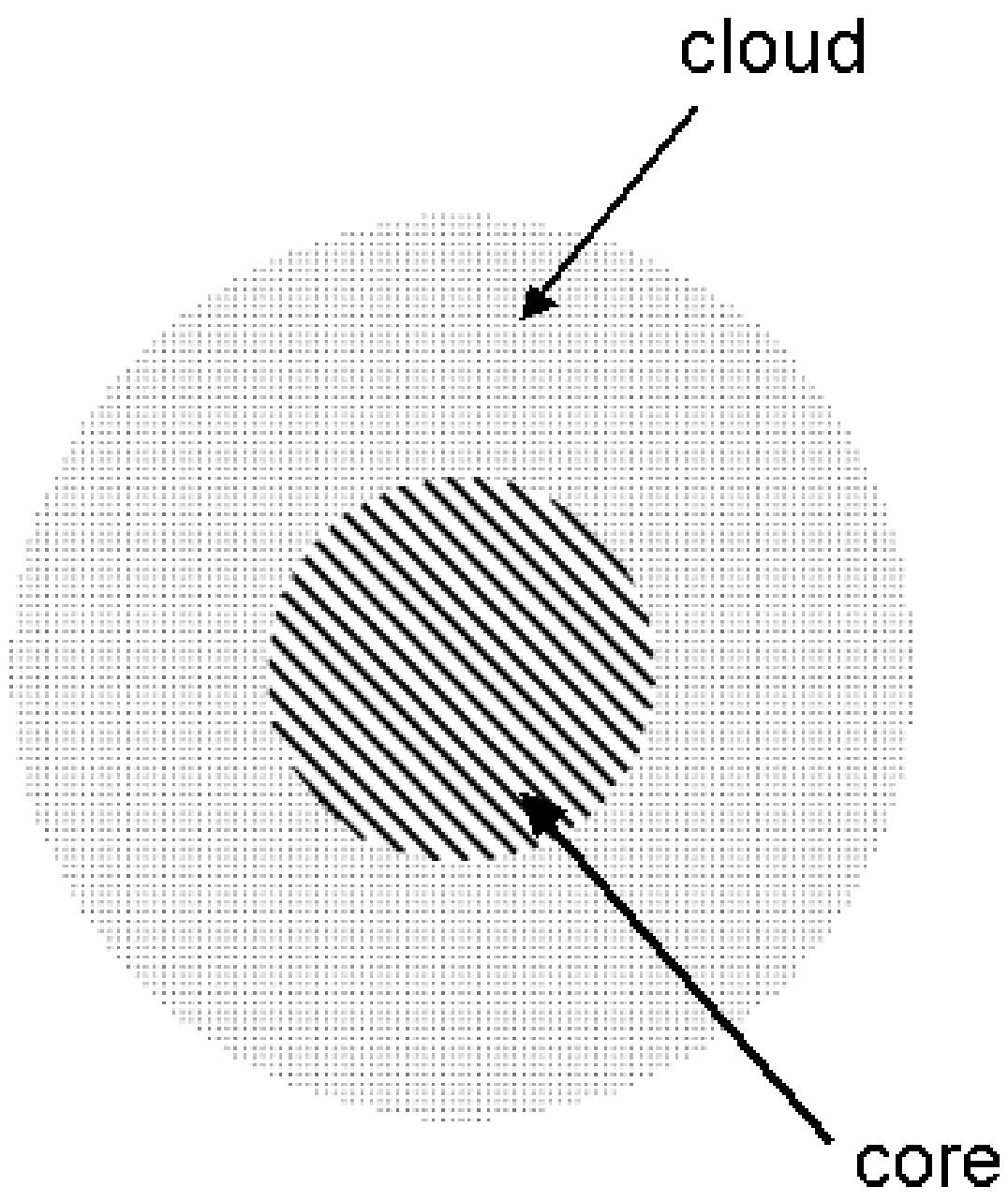

Fig. 1. Initial description of the nucleon having an outer cloud and an inner core.

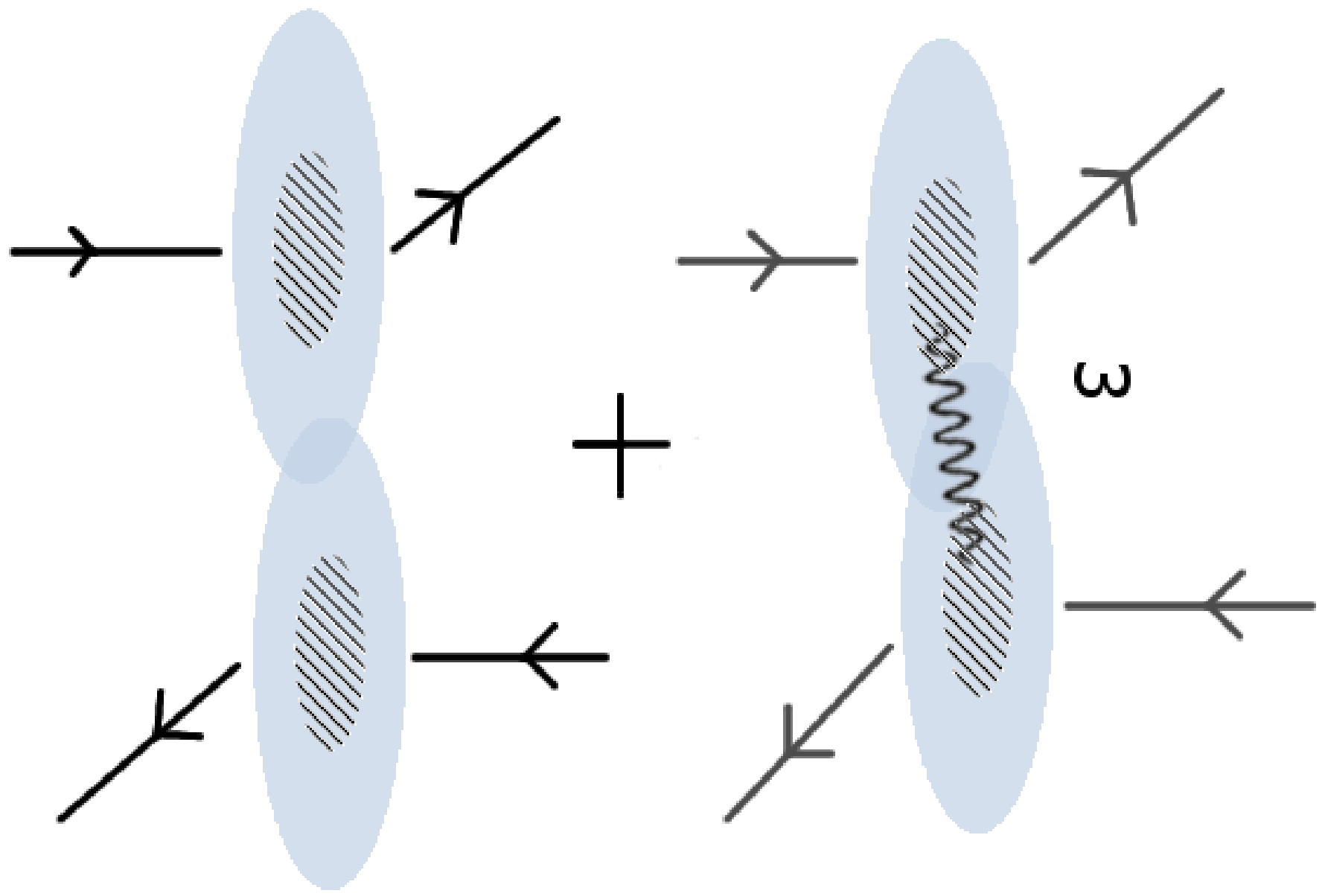

a) b)

Fig. 2. High energy elastic scattering viewed: a) as a glancing collision between nucleon clouds; b) as a hard or large $|t|$ collision between nucleon cores via $\omega$ exchange with their clouds overlapping.

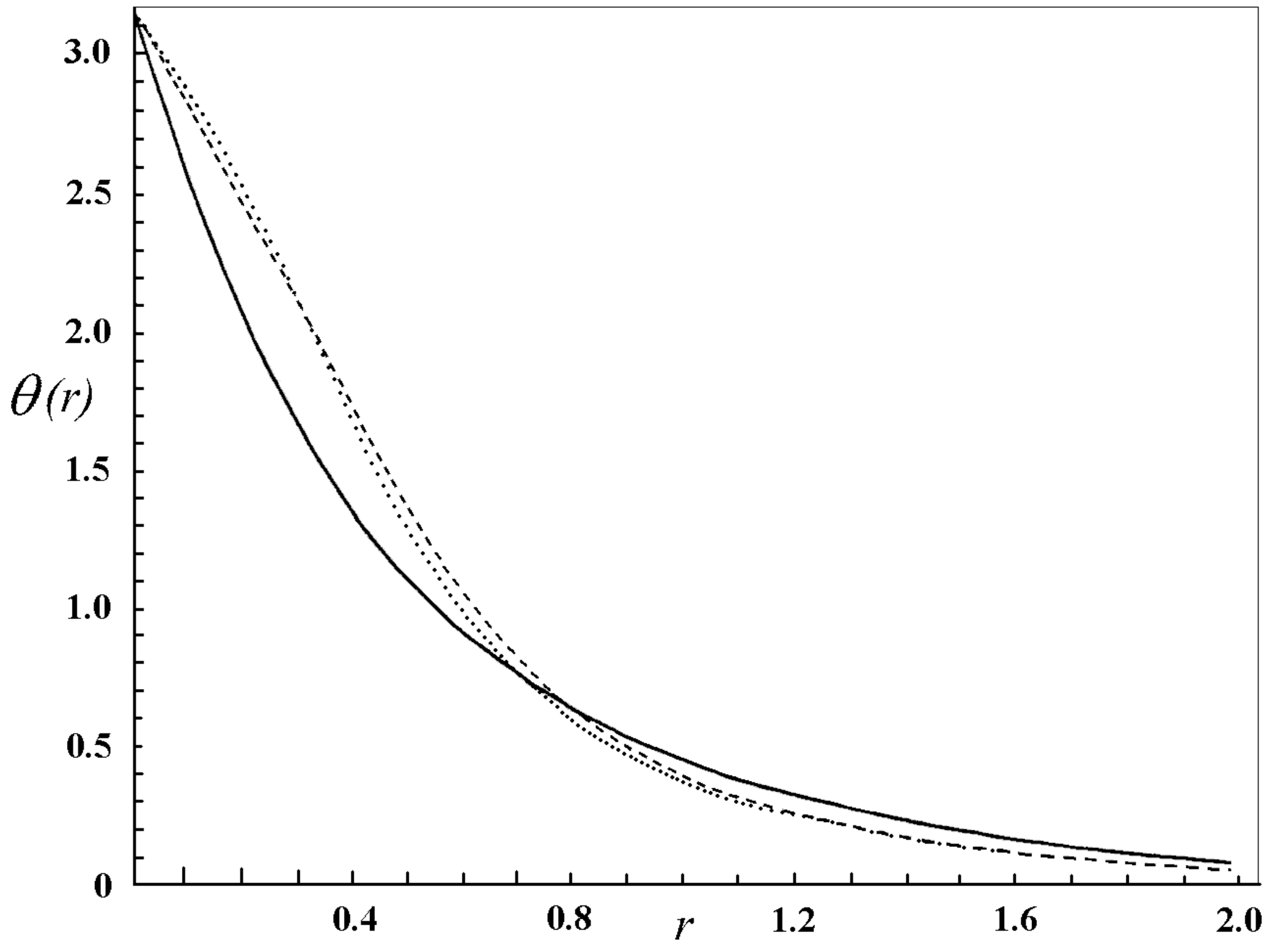

Fig. 3. The pion profile function $\theta(r)$ as a function of $r$ in Fermi. The solid curve represents $\theta(r)$ obtained from high energy elastic scattering. The dotted and the dashed curves represent $\theta(r)$ obtained from low energy calculations in the minimal and in the complete soliton model by Meissner et al. (ref. 28).

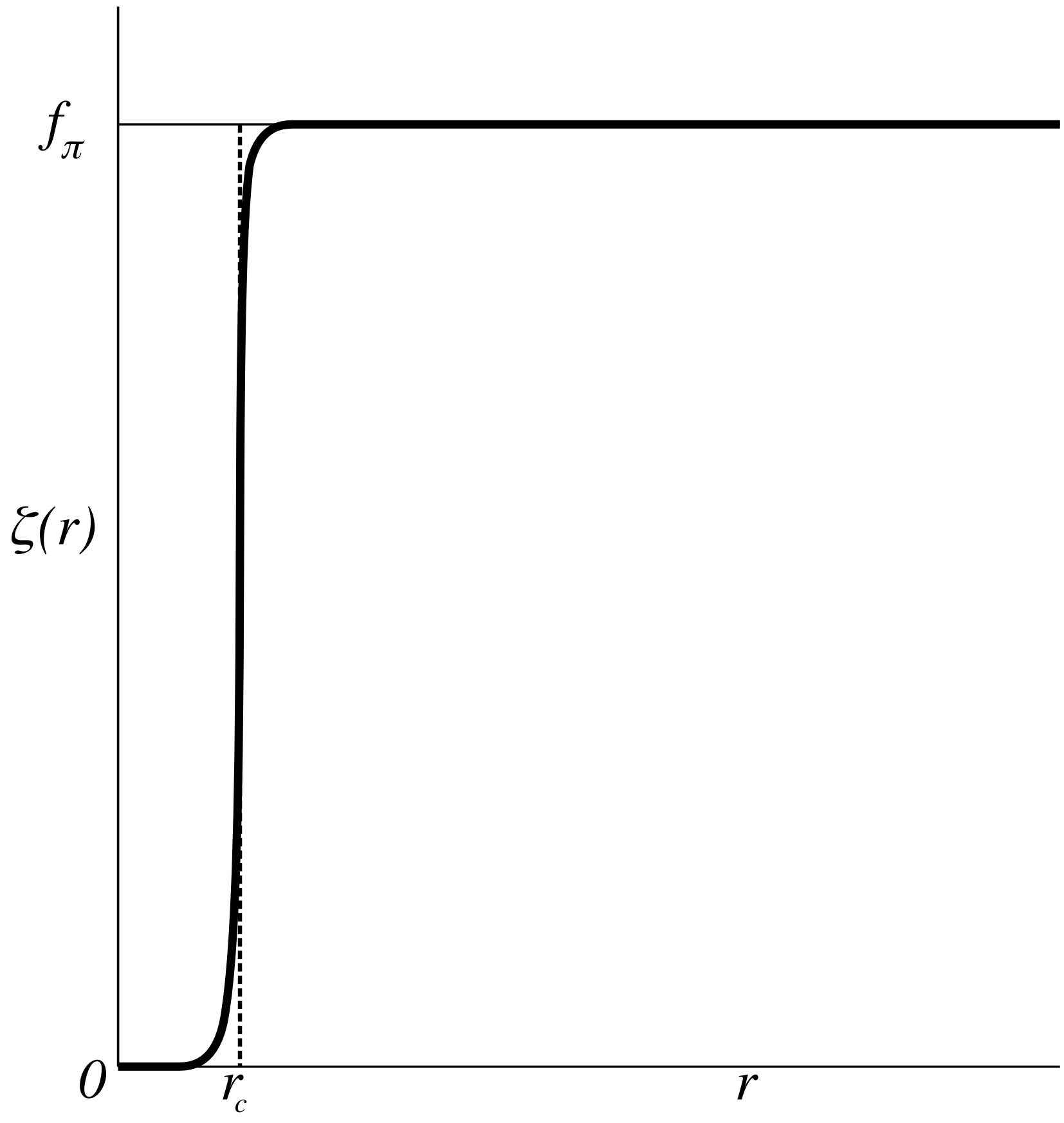

Fig. 4. Critical behavior of the scalar field $\zeta(r)$ as a function of $r$ : $\zeta(r) = 0$ for $r$ less than a critical distance $r_c$ and rises sharply at $r = r_c$ to its vacuum value $f_\pi$ .

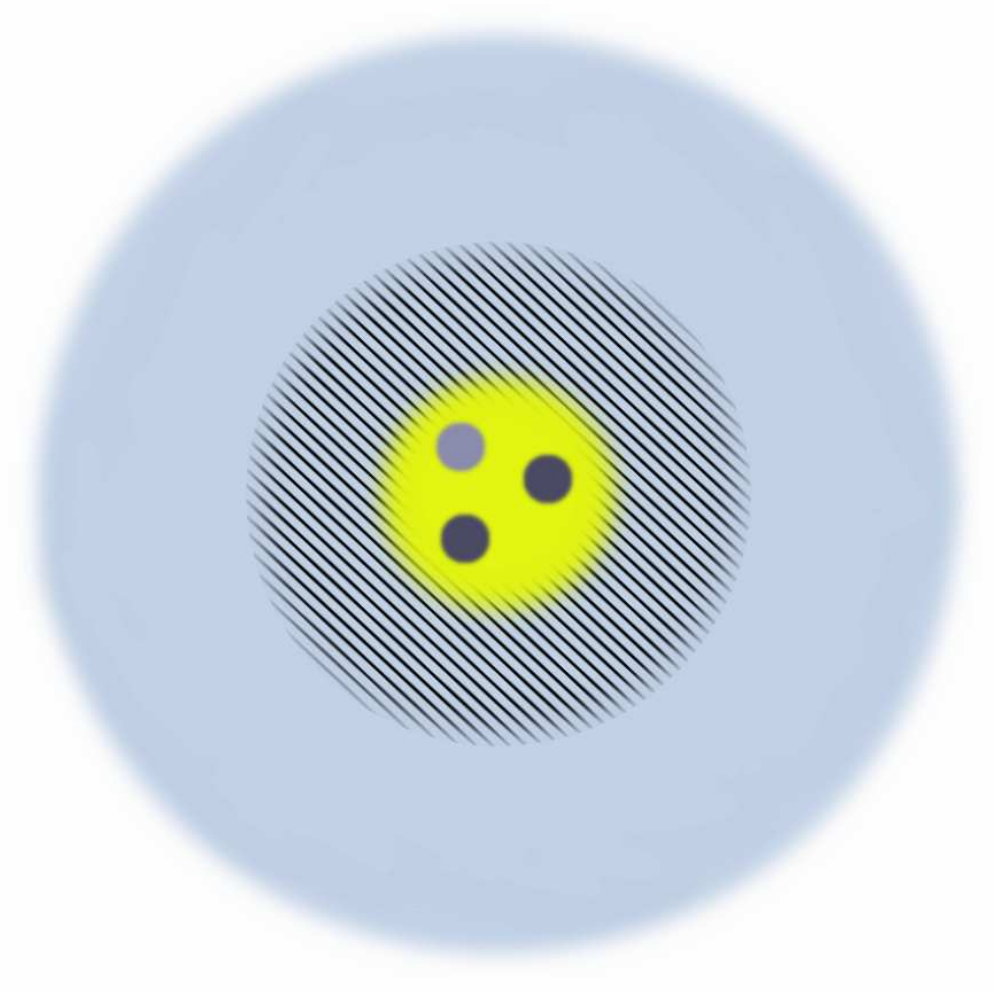

Fig. 5. Nucleon structure emerging from our investigation. Nucleon has an outer cloud of $q\bar{q}$ condensed ground state analogous to the BCS ground state in superconductivity, an inner core of topological baryonic charge probed by $\omega$, and a still smaller quark-bag of massless valence quarks.

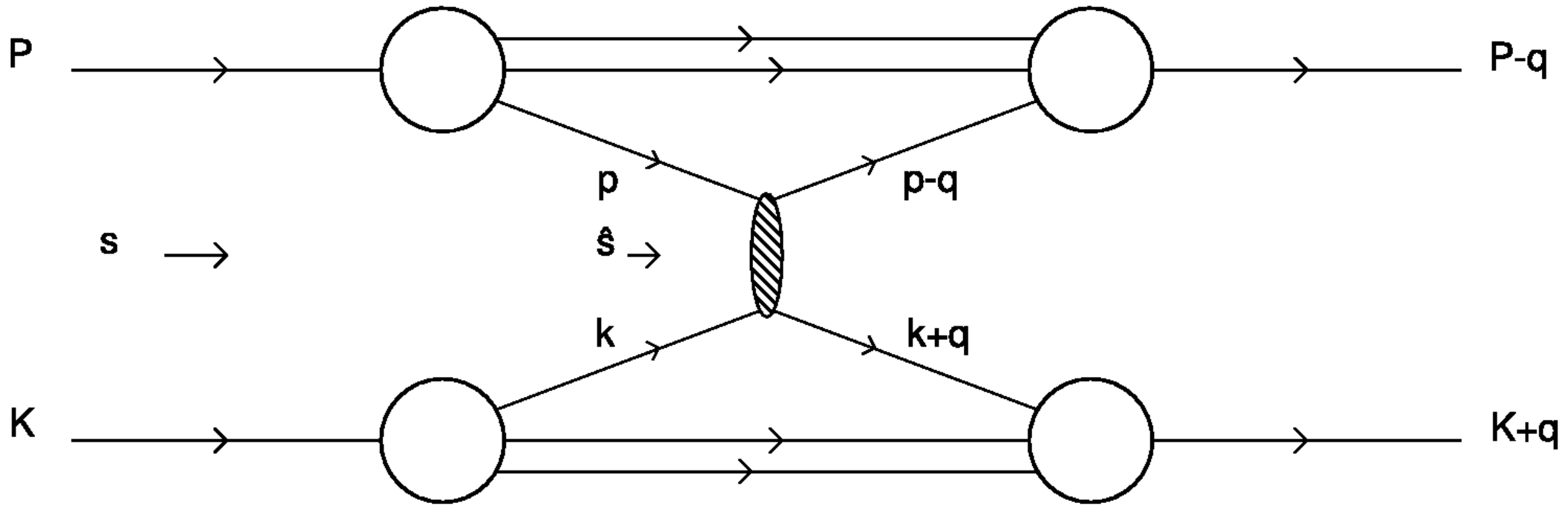

Fig. 6. Hard collision of valence quarks from two colliding protons.

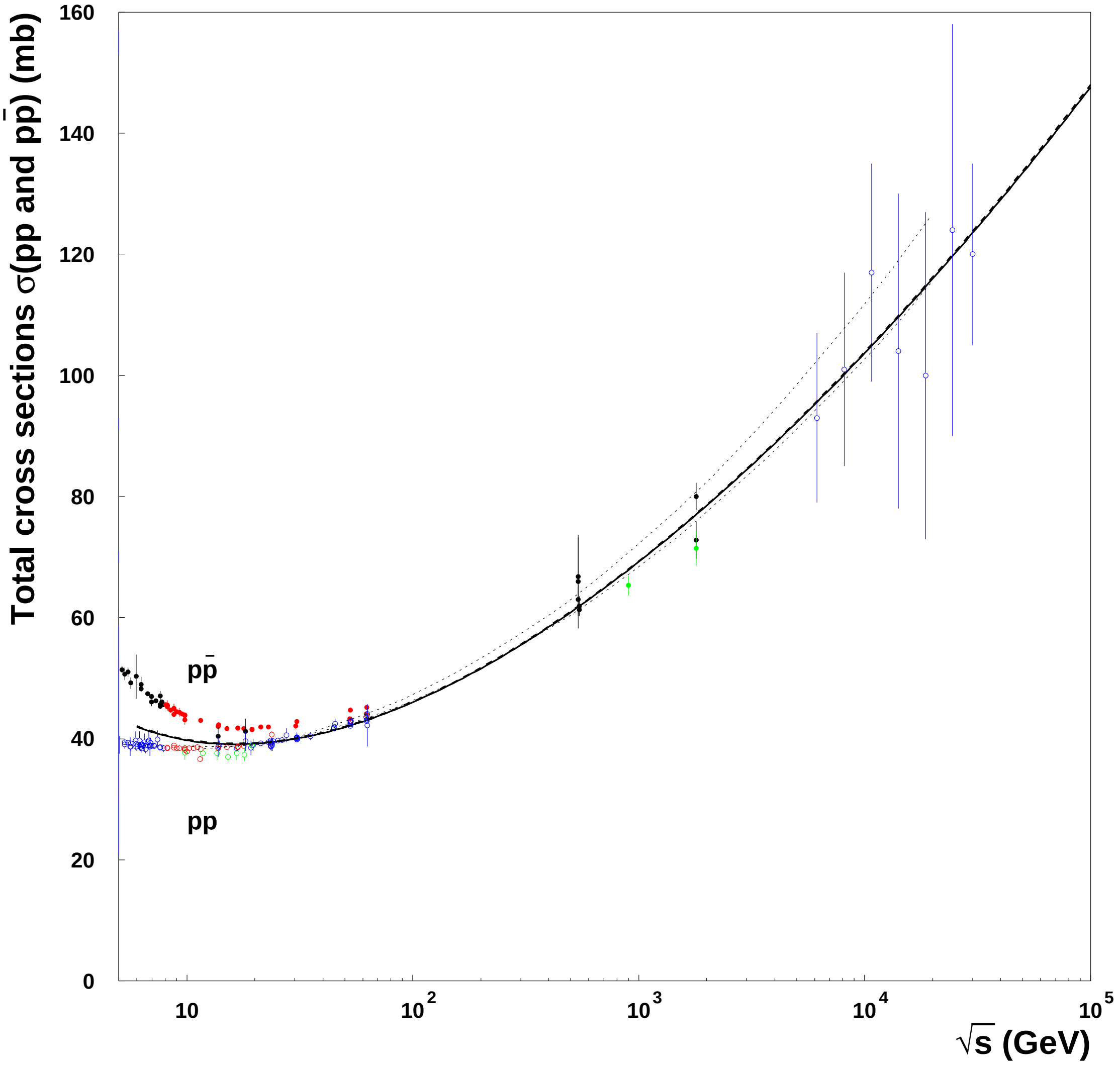

Fig. 7. Solid curve represents our calculated total cross section as a function of $\sqrt{s}$ . Dotted curves represent the error band given by Cudell et al. (ref. 53).

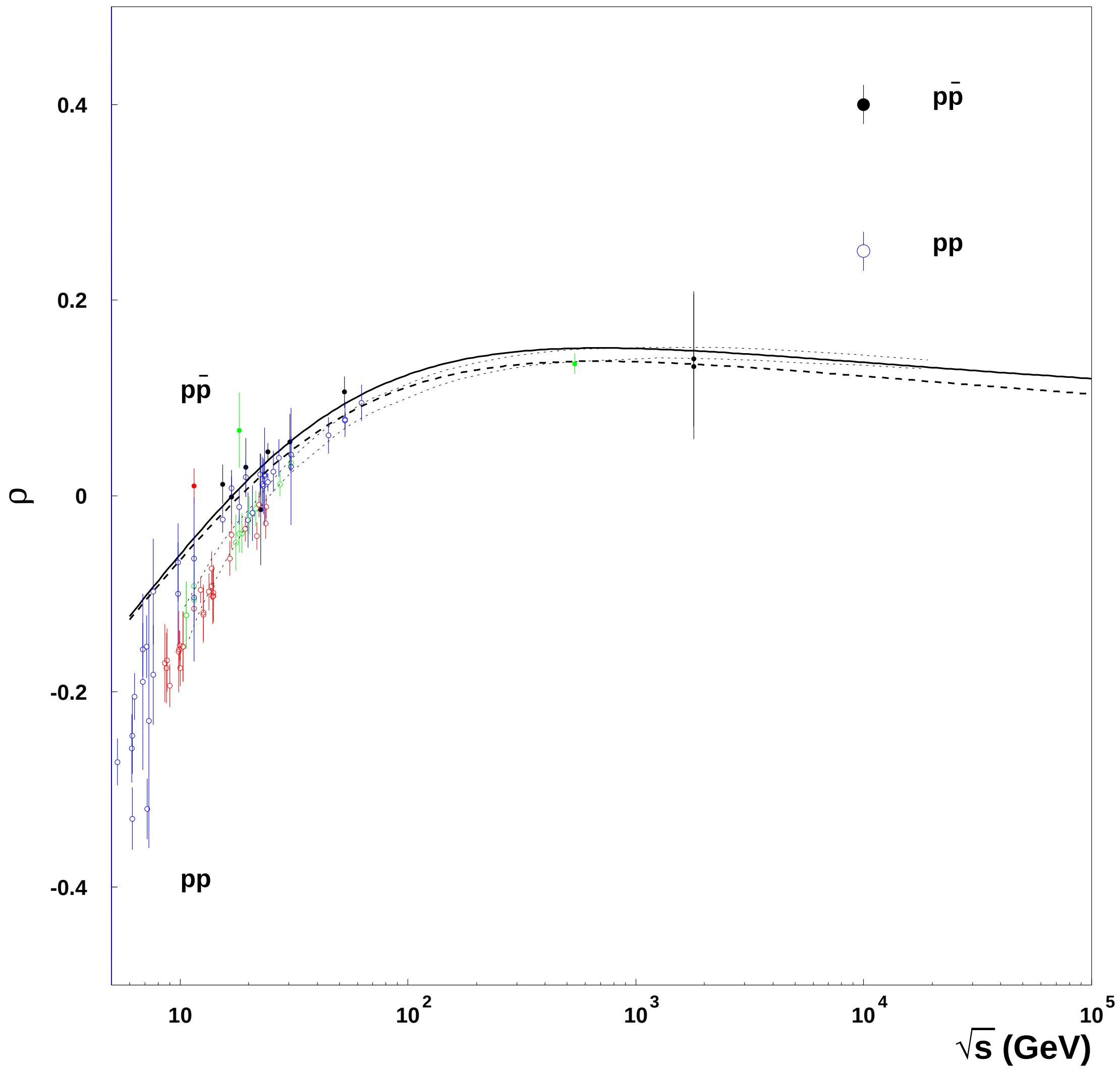

Fig. 8. Solid and dashed curves represent our calculated $\rho_{\bar{p}p}$ and $\rho_{pp}$ as functions of $\sqrt{s}$. Dotted curves represent the error band given by Cudell et al. (ref. 53).

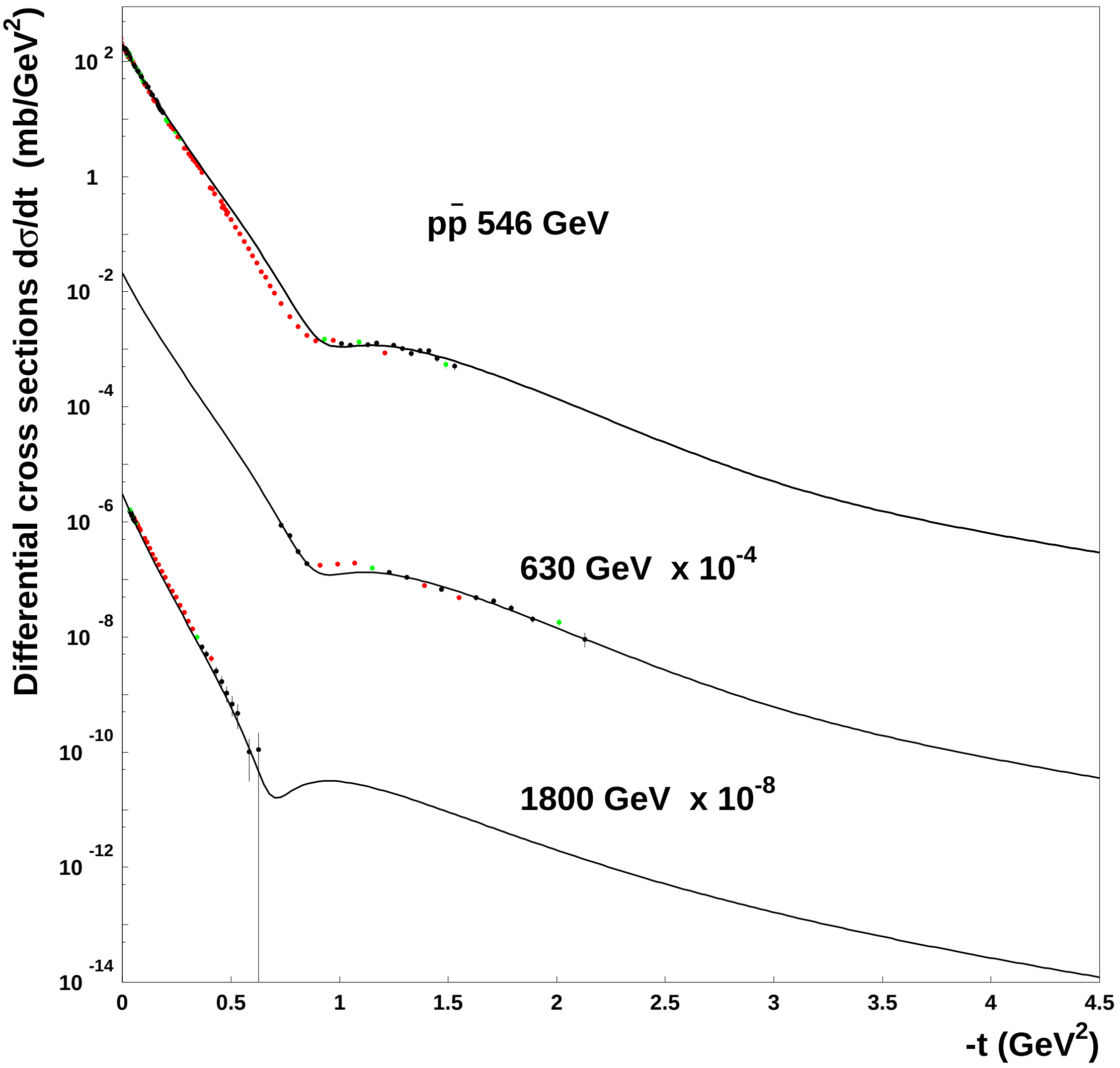

Fig. 9. Solid curves show our calculated $d\sigma/dt$ at $\sqrt{s}$ =546, 630 and 1800 GeV. Experimental data are from refs. (2, 3 and 5, 6).

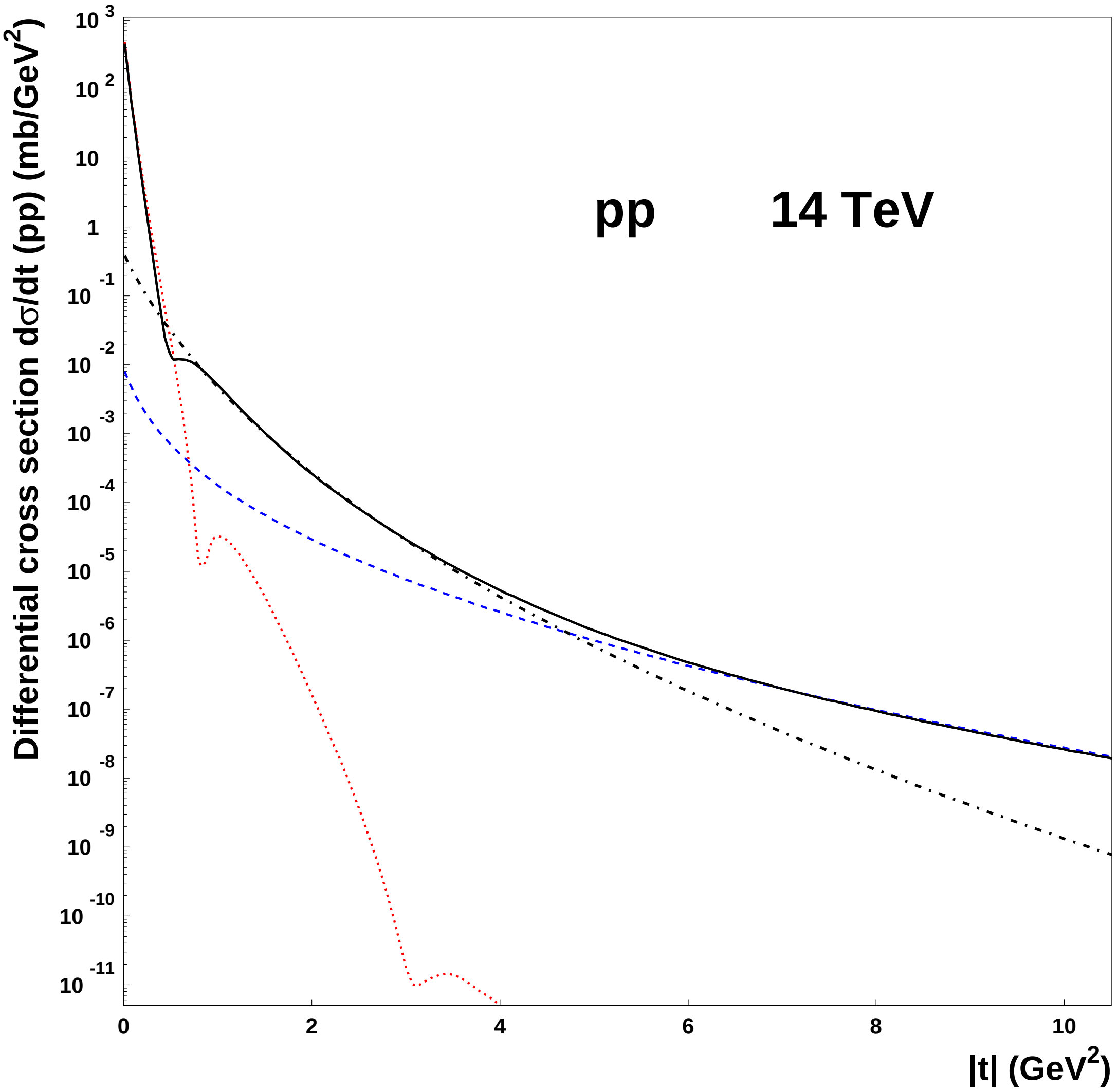

Fig. 10. Solid curve shows our predicted $d\sigma/dt$ for $pp$ elastic scattering at LHC energy $\sqrt{s}$ =14 TeV (ref. 14). Dotted curve represents $d\sigma/dt$ due to diffraction only. Similarly, dot-dashed curve and dashed curve represent $d\sigma/dt$ due to hard $\omega$-exchange and hard $qq$ scattering only.

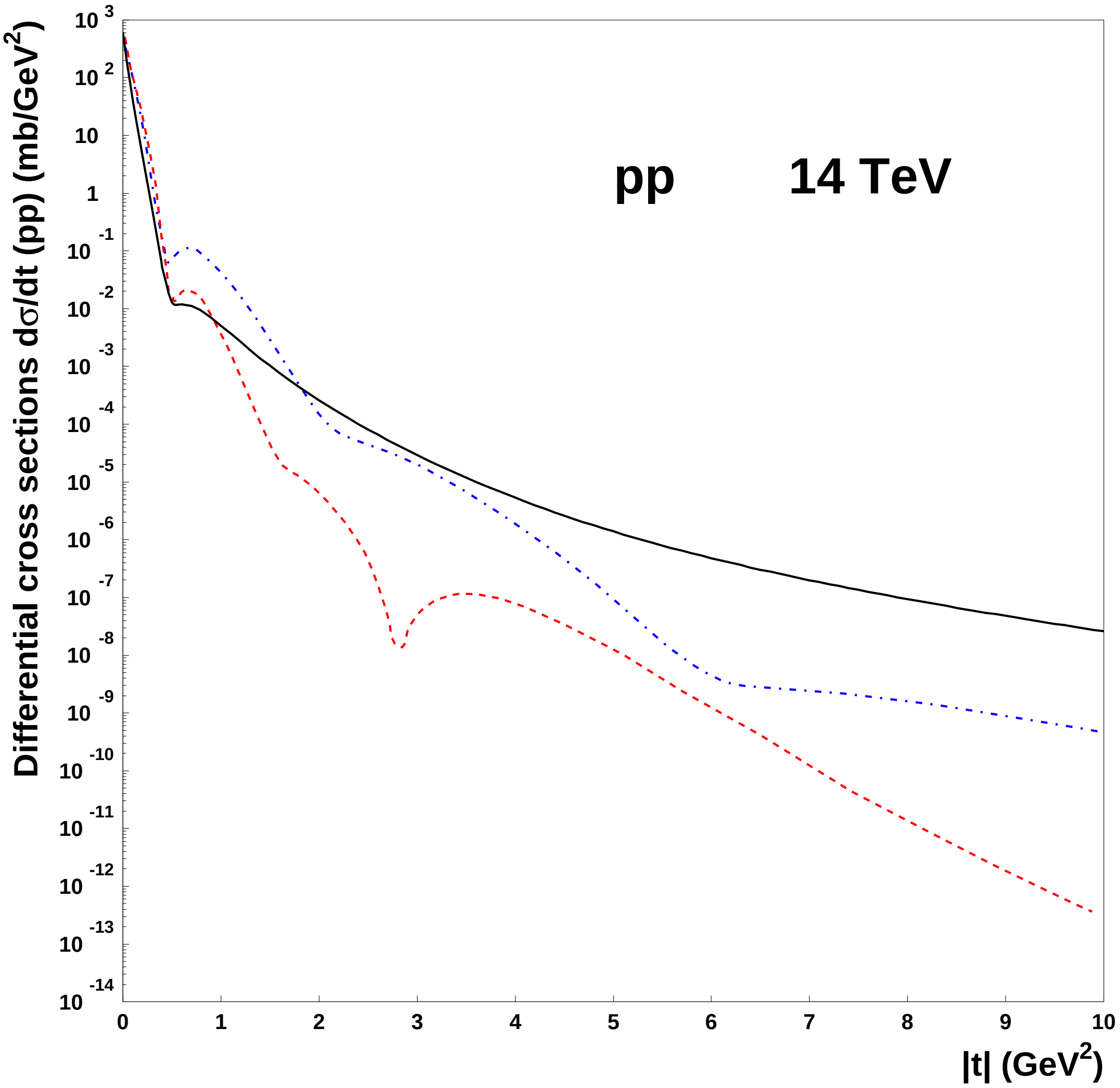

Fig. 11. Comparison of our predicted $pp$ elastic $d\sigma/dt$ at LHC energy $\sqrt{s}$ =14 TeV (solid curve) with those of Bourrely et al. (dot-dashed curve) and Desgrolard et al. (dashed curve). We predict for $|t| \gtrsim 1$ GeV$^2$ a smooth fall of $d\sigma/dt$, while they predict oscillations. Also, our predicted $d\sigma/dt$ at large $|t|$ is much higher than theirs.

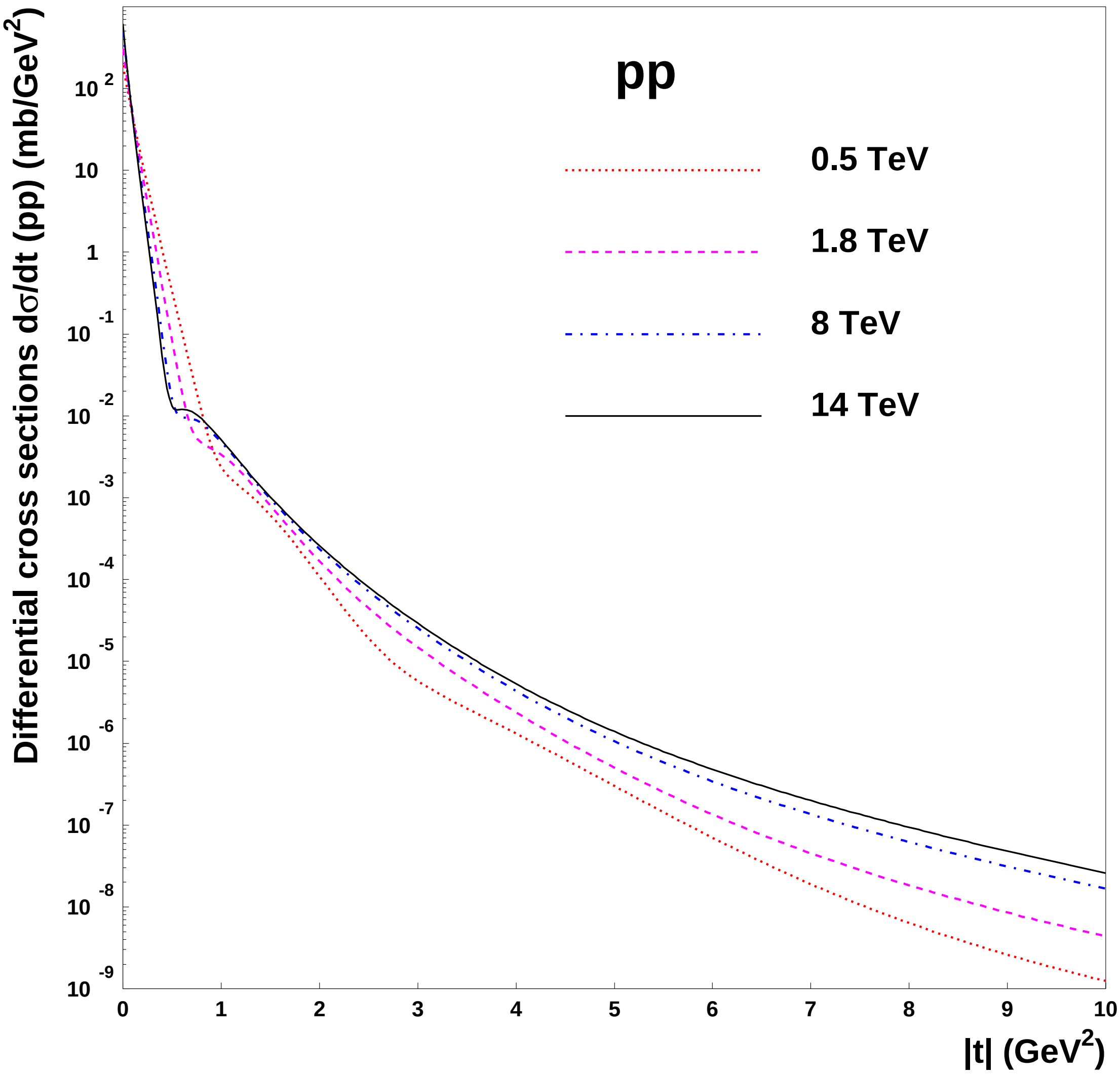

Fig. 12. $pp$ elastic $d\sigma/dt$ at different energies in the $|t|$ range: $|t| = 0 - 10$ GeV$^2$. The curves illustrate the energy dependence of the differential cross section. They also show that $d\sigma/dt$ at large $|t|$ increases with energy (because of the hard pomeron).

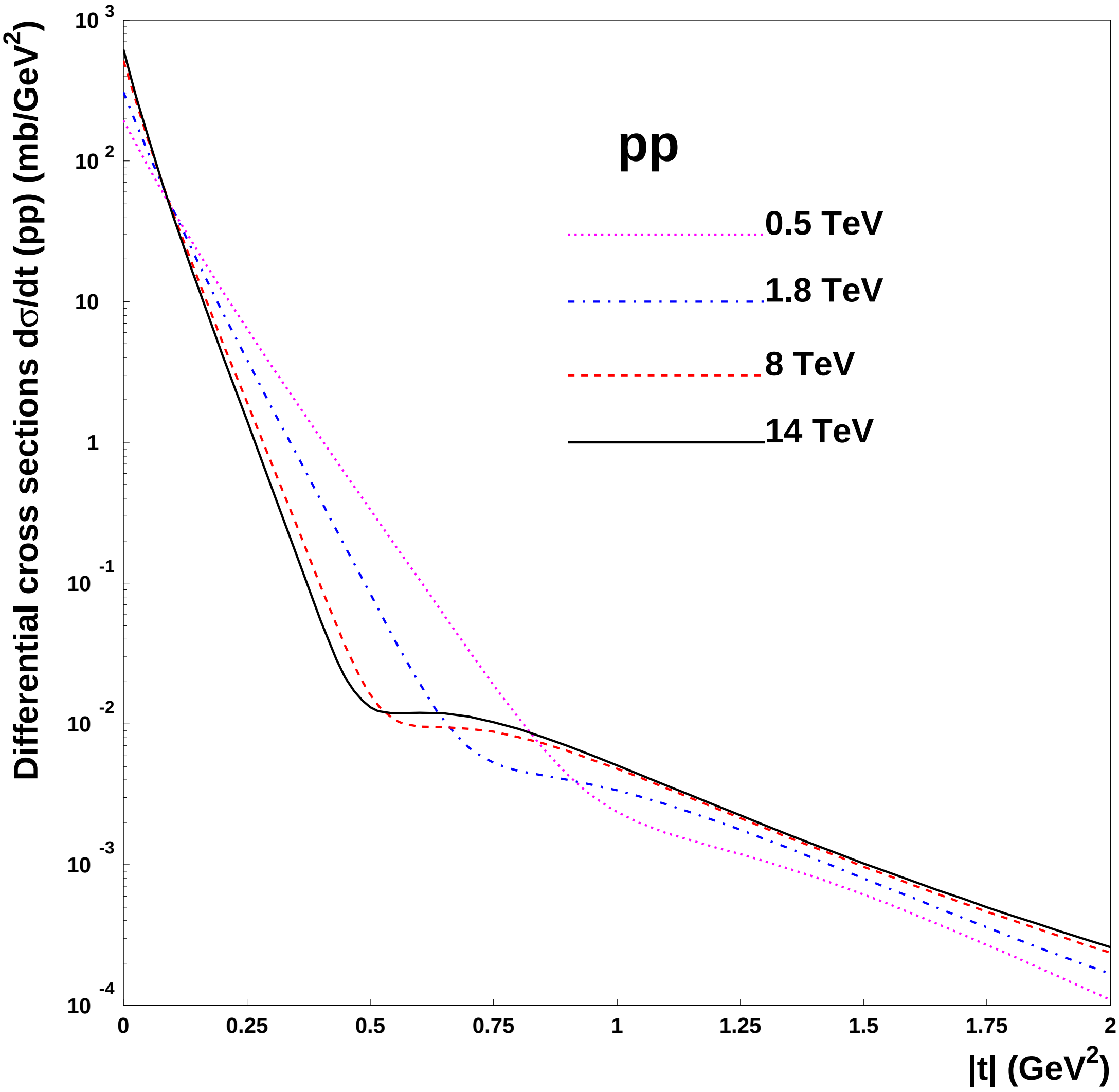

Fig. 13. Same as Fig. 12 for a smaller $|t|$ range: $|t| = 0 - 2\,\mathrm{GeV}^2$. These curves clearly show the rise in $d\sigma/dt$ at $|t| = 0$, the shrinkage of the diffraction peak, and the movement of the shoulder toward smaller $|t|$ with increasing energy.

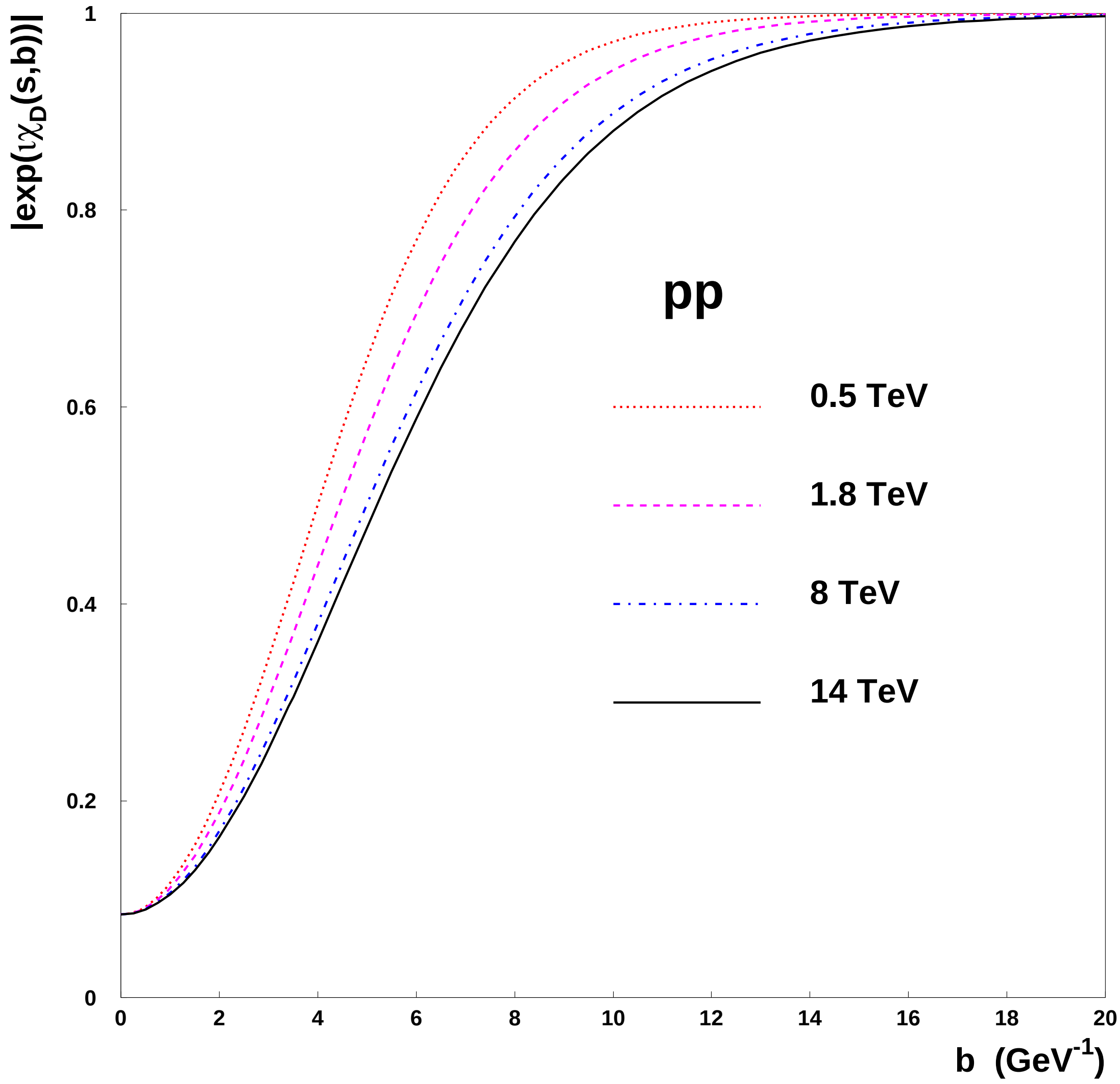

Fig. 14. $\left|\exp[i\chi_D^+(s,b)]\right|$ as a function of $b$ at different energies showing that unitarity is always satisfied. The curves also show that for $b=0$, $\left|\exp[i\chi_D^+(s,0)]\right|$ is essentially energy independent and finite.

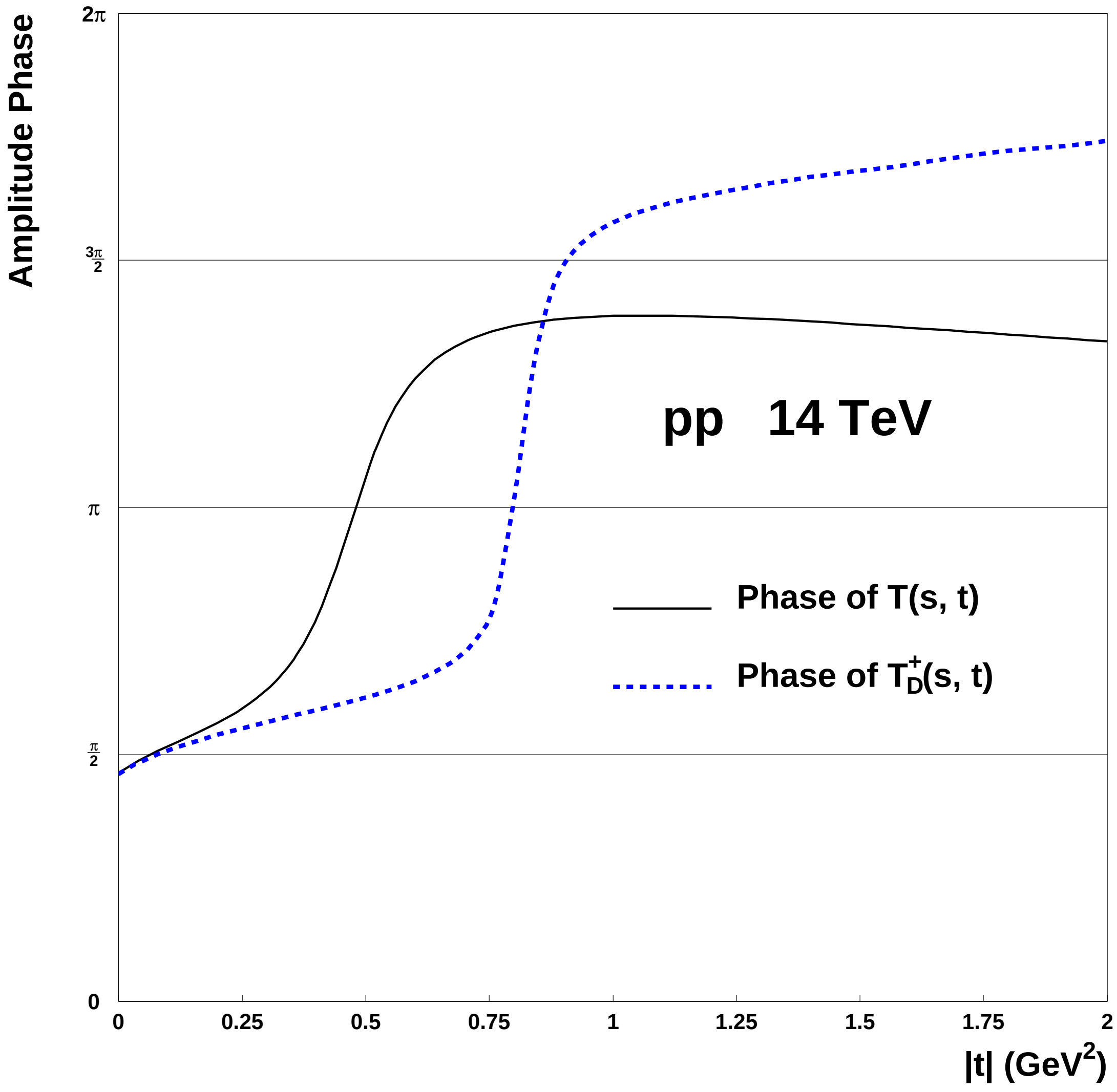

Fig. 15. Phases of the diffraction amplitude $T_D^+(s,t)$ (dashed curve) and the full amplitude $T(s,t)$ (solid curve) as functions of $|t|$ at $\sqrt{s}$ = 14 TeV.

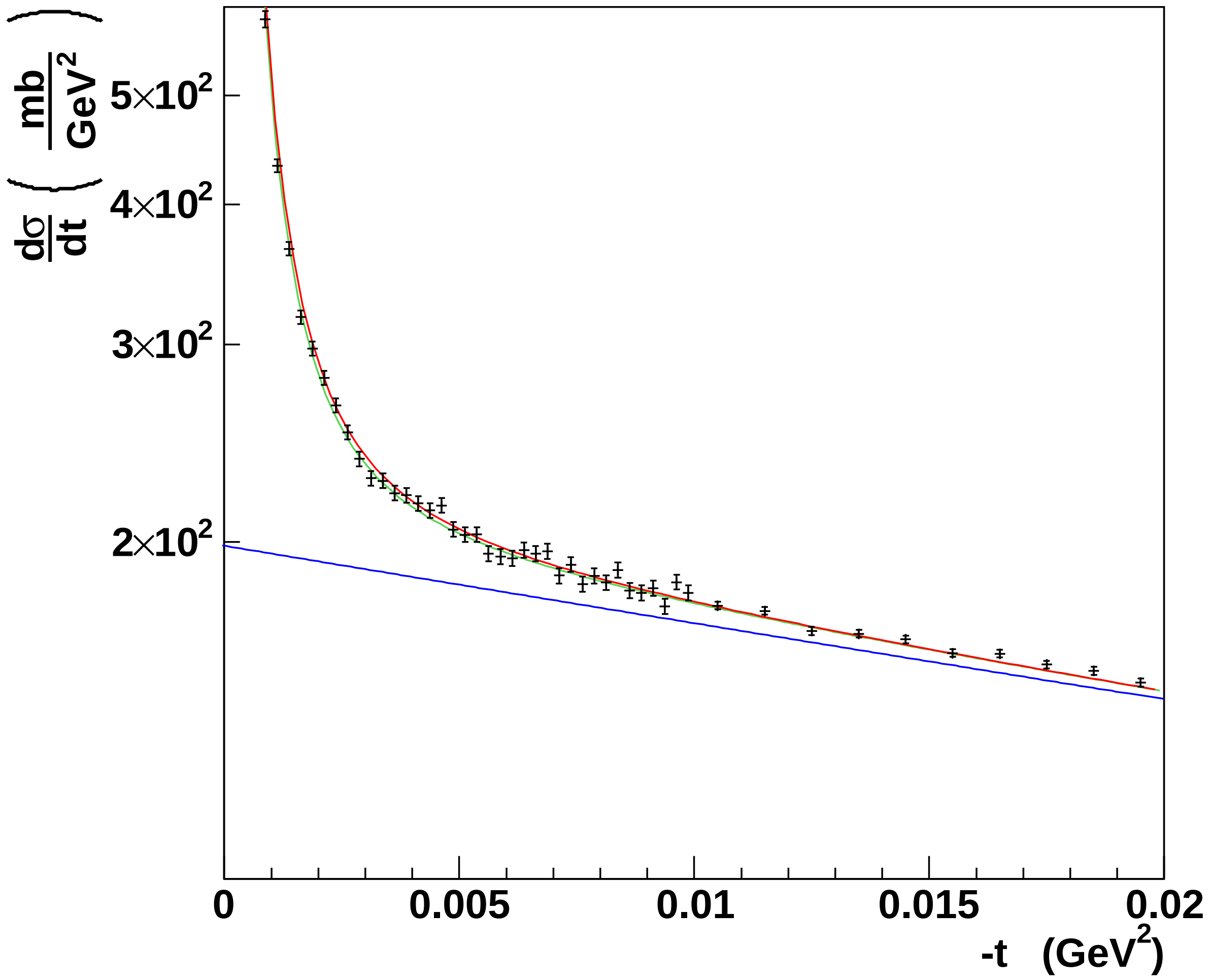

Fig. 16. $d\sigma/dt$ for $\bar{p}p$ elastic scattering at $\sqrt{s}$ = 541 GeV when Coulomb interaction is included in our model (Kaspar, ref. 54). The two rising curves have been calculated by Kaspar using for the combined Coulomb-hadronic amplitude the West-Yennie formulation (lower curve) and the Kundrat-Lokajicek formulation (upper curve). The straight line represents $d\sigma/dt$ due to our hadronic amplitude alone. Experimental data are from Augier et al. (ref. 55).

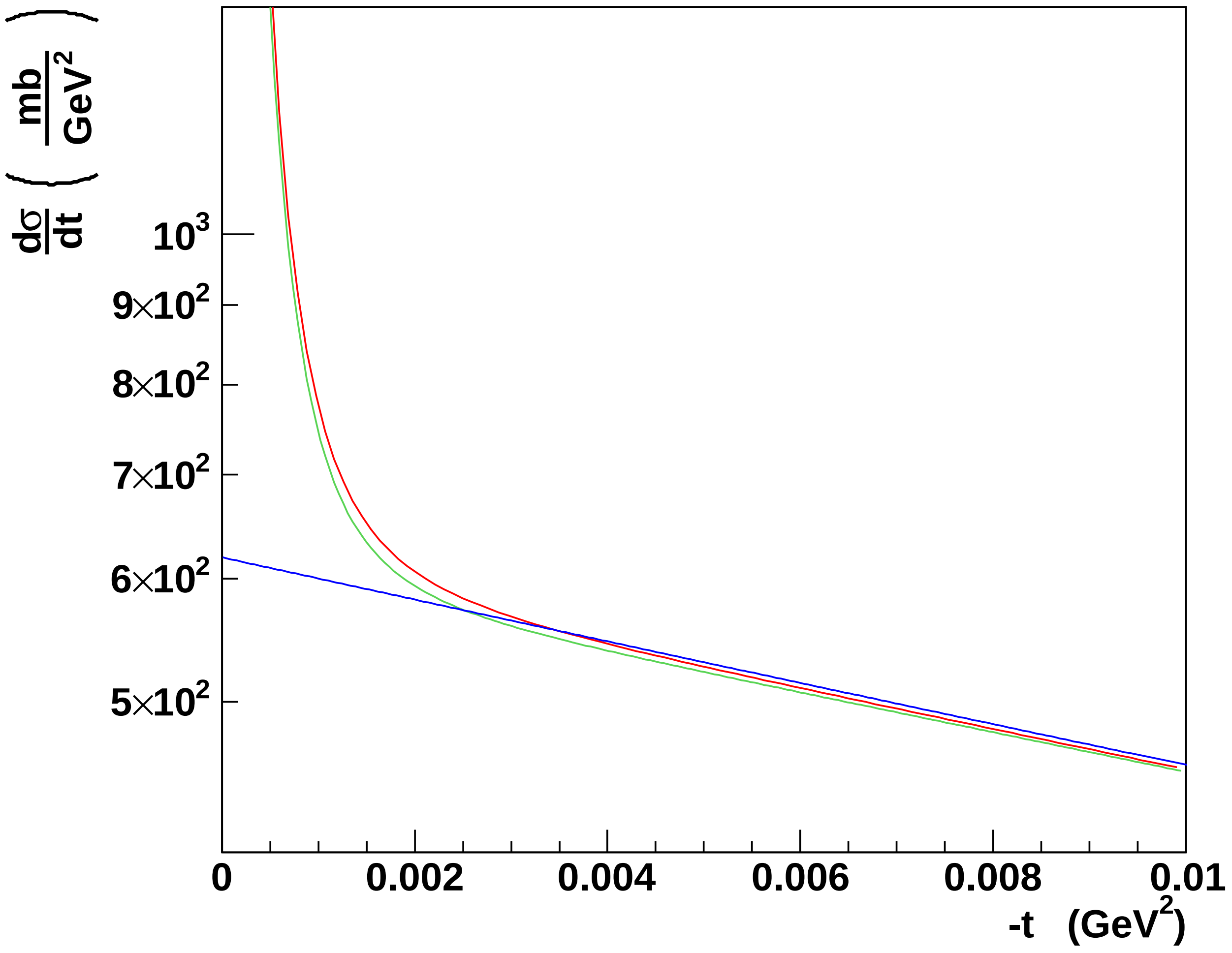

Fig. 17. Same calculations by Kaspar as in Fig. 16 for *pp* elastic scattering at $\sqrt{s}$ =14 TeV.